\documentclass[journal]{IEEEtran}

\usepackage{graphicx}
\usepackage{soul, color}
\usepackage{tabularx}
\usepackage{adjustbox}
\newcommand{\comm}[1]{}
\usepackage{longtable}
\usepackage{comment}
\usepackage{hyperref}
\usepackage{amsmath}
\usepackage{color}
\usepackage{array}
\usepackage{xtab}
\usepackage{cite}
\usepackage{stfloats}
\usepackage{multicol}
\usepackage{bm}
\usepackage{amssymb}
\title{Adaptive Anomaly Detection for Identifying Attacks in Cyber-Physical Systems: A Systematic Literature Review}

\author{
    Pablo Moriano$^{\ast}$,~\IEEEmembership{Senior Member,~IEEE}, 
    Steven C. Hespeler$^{\ast}$, 
    Mingyan Li, 
    and Maria Mahbub~\IEEEmembership{Member,~IEEE}
    \thanks{$^{\ast}$Pablo Moriano and Steven C. Hespeler contributed equally to this work and are considered co-primary authors.}
    \thanks{Pablo Moriano and Steven C. Hespeler are with the Computer Science and Mathematics Division, Oak Ridge National Laboratory, Oak Ridge, TN, USA (e-mail: \{moriano@ornl.gov, hespelersc@ornl.gov\}).}%
    \thanks{Mingyan Li and Maria Mahbub are with the Cyber Resilience and Intelligence Division, Oak Ridge National Laboratory, Oak Ridge, TN, USA (e-mail: \{lim3@ornl.gov, mahbubm@ornl.gov\}).}%
    \thanks{This manuscript has been co-authored by UT-Battelle, LLC, under contract DE-AC05-00OR22725 with the US Department of Energy (DOE). The US government retains and the publisher, by accepting the article for publication, acknowledges that the US government retains a nonexclusive, paid-up, irrevocable, worldwide license to publish or reproduce the published form of this manuscript, or allow others to do so, for US government purposes. DOE will provide public access to these results of federally sponsored research in accordance with the DOE Public Access Plan (\url{http://energy.gov/downloads/doe-public-access-plan}).}%
}

\begin{document}

\maketitle

\begin{abstract} \label{sec:abstract}
Modern cyberattacks in cyber-physical systems (CPS) rapidly evolve and cannot be deterred effectively with most current methods which focused on characterizing past threats. Adaptive anomaly detection (AAD) is among the most promising techniques to detect evolving cyberattacks focused on fast data processing and model adaptation. AAD has been researched in the literature extensively; however, to the best of our knowledge, our work is the first systematic literature review (SLR) on the current research within this field. We present a comprehensive SLR, gathering 397 relevant papers and systematically analyzing 65 of them (47 research and 18 survey papers) on AAD in CPS studies from 2013 to 2023 (November). We introduce a novel taxonomy considering attack types, CPS application, learning paradigm, data management, and algorithms. Our analysis indicates, among other findings, that reviewed works focused on a single aspect of adaptation (either data processing or model adaptation) but rarely in both at the same time. We aim to help researchers to advance the state of the art and help practitioners to become familiar with recent progress in this field. We identify the limitations of the state of the art and provide recommendations for future research directions.
\end{abstract}
\begin{IEEEkeywords}
Cyber-Physical Systems, Anomaly Detection, Adaptive Learning, Cybersecurity.
\end{IEEEkeywords}
\section{Introduction} \label{sec:intro}

\IEEEPARstart{M}{odern} cyber-physical systems (CPS) including industrial control systems (ICS), vehicles, power grids, and the Internet of Things (IoT), among others, generate vast amounts of high speed data that need to be processed to support decision-making capabilities~\cite{Atat:2018:Big:Data:CPS:Panoramic:Survey, Fei:2019:CPS:Stream:Analytics, Kayan:2022:Cybersecurity:Industrial:CPS}. Due to the mission-critical nature that CPS play, securing its operation against malicious threats is essential for guaranteeing daily life activities. However, since information technology and physical processes are closely linked in CPS, they face a broader range of threats, including both cyber and physical attacks. Here, cyberattacks refer to attacks against the communication and computing components of CPS~\cite{Shacham:2004:Effectiveness:Address:Space:Randomization, Pike:2016:Trackos:Security:Aware:Real:Time:OS, Clements:2017:Protecting:embedded:Systems:Privilige:Overlays}. Physical attacks refers to compromising the physical environment of a CPS, subjecting the system to potential malicious data via injection or tampering through sensor and/or actuators~\cite{Rutkin:2013:GPS:Spoofers:Yacth, Shoukry:2013:Non:Invasive:Spoofing:Attacks:Anti:Lock:Break, Petit:2015:Remote:Attacks:Vehicle:Sensors:Camera:Lidar}.

Among the most used techniques to detect threats in CPS are those based on anomaly detection~\cite{Chandola:2009:Anomaly:Detection:Survey, Mirsky:2018:Kitsune, Moriano:2021:Burstiness:BGP, Moriano:2022:AHC}. As opposed to traditional signature-based detection focused on matching patterns from previously seen attacks~\cite{Hubballi:2014:False:Alarm:Minimization:Techniques:Signature:IDS, Ioulianou:2018:Signature:Based:IDS:IoT, Wu:2019:Survey:Intursion:Detection:Vehicle:Networks}, anomaly detection techniques focus on spotting behavior that looks different from the expected norm~\cite{Luo:2021:DL:Anomaly:Detection:CPS, Shahriar:2023:CANshield, Moriano:2024:Benchmarking:Unsupervised:Online:IDS}. This approach helps on identifying previously unseen attacks as those commonly affecting the cyber and physical components of CPS~\cite{Schneider:2018:High:Performance:Unsupervised:Anomaly:Detection:CPS, Xi:2022:Data:Correlation:Aware:Unsupervised:DL:Anomaly:Detection}. Different anomaly detection approaches have been proposed to secure CPS including those based on attack-resilient sensor fusion~\cite{Ivanov:2016:Attack:Resilient:Sensor:Fusion:CPS, Lu:2018:Attack:Resilient:Sesnor:Fusion:Adaptive:Cruise:Control}, model-based attack detection~\cite{Quinonez:2020:SAVIOR:Securing:Autonomous:Vehicles:Robust:Physical:Invariants, Giraldo:2018:Survey:Physics:Based:Attack:CPS}, and data-based detection~\cite{Junejo:2016:Behaviour:attack:Detection:CPS:ML, Shin:2017:Intelligent:Sensor:Attack:Detection:Automobile}. However, the ability of anomaly detection methods to adapt to detect previously unseen attacks is usually not thoroughly explored. Here, adaptability is closely tied with the concept of ``adaptation,'' which means the ability of an anomaly detection method to anticipate and respond to new and emerging security threats by learning from the experience and the current state of the CPS~\cite{Abie:2019:Cognitive:Cyber:CPS:IoT, Andrade:2019:Cognitive:Security}.

Adaptive anomaly detection (AAD) requires two key components: (1) near real-time data processing and (2) a predefined learning mode for model adaptation~\cite{Raciti:2013:Anomaly:And:Adaptation:CPS, Settanni:2018:Protecting:CPS:AD:Adaptation, Akowuah:2021:Adaptive:Sensor:Attack:Autonomous:CPS, Biggio:2024:ML:Security:Difficult:Fix}. Near real-time processing assists in detecting attacks before they cause consequences, which is crucial in safety-critical CPS. In addition, a predefined learning model (such as full, incremental, or hybrid retrain~\cite{Gama:2014:Survey:Concept:Drift}) is needed to adapt the detection model to respond better to unseen attacks. Both components of AAD generally ensure a strong defense against advanced cyberattacks.

This paper presents a systematic literature review (SLR) on AAD in CPS. Our goal is to provide a comprehensive overview of the state of the art in AAD, including their usage across different types of CPS, classification across different learning paradigms, common algorithms, and a discussion of trends and gaps in the literature. To our knowledge, this is the first SLR on AAD in CPS. More specifically, the objectives of our SLR are to: (1) provide researchers with an understanding of current AAD methods in CPS, enabling new researchers to quickly familiarize themselves; (2) highlight gaps and opportunities for future research; and (3) support practitioners in selecting and adapting AAD methods in CPS to fit their needs.

In the past decade, a large number of studies were published covering different aspects of AAD in CPS~\cite{Li:2016:Dirichlet:Based:Detection:Smart:Grid, Adhikari:2017:Applying:Adaptive:Threes:Real:Time:IDS, Van:2019:Real:Time:Anomaly:Detection:Automated:Vehicles, Yasaei:2020:IoT:CAD:Context:Aware:Anomaly:Detection:Through:Sensor:Association, Mowla:2020:AFRL:Adaptive:Federated:RL:Jamming, Jiao:2022:Cyberattack:Resilient:Forecasting:Adaptive:Robust:Regression, Ding:2022:Data:Driven:Situational:Awarness:Power:systems, Gyamfi:2022:Novel:Online:IDS:IoT:OI-SVDD:AS-ELM, intriago2023real, cai2023adam}. A small amount of these surveys have also explored the application of machine learning and data mining techniques to various cybersecurity domains, with an explicit focus on addressing intrusion detection challenges for securing CPS and providing insights into methodologies and best practices~\cite{buczak2015survey, olowononi2020resilient}. Nonetheless, most studies investigated focus on only one of the key components of AAD (i.e., near real-time data processing or predefined learning mode). Accordingly, results were contradictory and practices heterogeneous. This makes it difficult to contextualize their contribution in terms of how adaptation is carried out.

To fill in the gaps and provide an updated and comprehensive review on the latest developments in AAD in CPS, the present study reviews state-of-the-art works published between 2013 and 2023 (November). The contributions of this article can be summarized as follows:
\begin{itemize}
    \item[(1)] We review and classify state-of-the-art AAD methods in CPS considering type of application, learning paradigm, data management strategy, and algorithms, along with a comprehensive summary tables (see Tables~\ref{table: Adaptive anomaly detection in ICS applications}-\ref{table: Adaptive anomaly detection in smart grids}).
    \item[(2)] We introduce a novel AAD taxonomy for CPS that focuses of attack types, applications, and ML algorithms to categorize reviewed works based on learning paradigm and algorithms.
    \item[(3)] We identify and discuss limitations of reviewed AAD approaches for securing CPS.
    \item[(4)] We discuss priority future areas of research in this field.  
\end{itemize}

We organize this SLR as follows. In Section~\ref{sec:related work}, we contextualize our SLR with respect to other surveys in closely-related areas. Section~\ref{sec:Methods} details the methodology we used to conduct the SLR. Section~\ref{sec:Results} introduces an AAD taxonomy for CPS and synthesize previous research based on the learning paradigm of the algorithms. In Section~\ref{sec:discussion}, we discuss our findings and potential future research directions. Finally, we provide a brief summary of this SLR in Section~\ref{sec:conclusion}.

\section{Background} \label{sec:background}

This section provides a summary of key concepts related to AAD in CPS. Common attack types to CPS and frequently used evaluation metrics in the detection task are also discussed in this section.  

\subsection{Cyber-Physical Systems (CPS)}

A CPS contains both a physical and cyber system \cite{duo2022survey}. These systems are co-designed and co-engineered from a variety of domains to be adaptive, flexible, and situationally aware \cite{li2019detection}. Edward A. Lee~\cite{lee2015past} states that the majority of CPS monitor and control the physical entities through feedback loops where physical processes affect computations. According to Lee, CPS combines deterministic models such as differential equations and digital logic with physical systems, which creates a a new engineering discipline that requires unique models and methods. Lee points out that the terms CPS and cyberspace can be attributed to the same root, ``cybernetics", a term originated by Norbert Wiener \cite{wiener2019cybernetics}. These complex systems represent the integration of computational algorithms and physical processes that are designed to interact with embedded computers and networks.

\subsection{Attacks on CPS}

Attacks on CPS can be classified into three main categories;  availability, integrity, and confidentiality \cite{duo2022survey}. Availability attacks aim to block or disrupt the communication network, making data and information unavailable to legitimate users. Integrity attacks target the accuracy and trustworthiness of data and control commands of a system. Confidentiality attacks focus on unauthorized access to sensitive information, we describe them in more detail below. 

\begin{itemize}
    \item Availability Attacks:
    \begin{itemize}
        \item \textit{Denial of Service (DoS)}: Prohibiting of the connection between communication. These attacks aim to make the communication services unavailable through means of bombarding the system with a large number of frames \cite{rajapaksha2023ai}. 
        \item \textit{Fuzzing}: Attacker sends a large number of arbitrary messages into the network(s) \cite{verma2024comprehensive}.
    \end{itemize}
    \item Integrity Attacks:
    \begin{itemize}
     \item \textit{False Data Injection (FDI)}: Insertion of false data into a system to disrupt operations or cause incorrect responses \cite{duo2022survey}. Duo et al. state that a false data injection can be identified by measuring the residuals and comparing them to a predefined threshold. 
     \item \textit{Masquerade or Impersonation}: An advanced form of attack, entails an attacker halting messages from a specific signal sent by a compromised node \cite{verma2024comprehensive}. In this situation, an attacker uses a fully compromised node to send fake messages at convincing frequency to imitate the target node \cite{verma2024comprehensive, rajapaksha2023ai}.
     \item \textit{Spoofing}: Malicious messages are injected into network \cite{rajapaksha2023ai}.
     \item \textit{Replay}: Upon hijacking a system, attackers observe and record readings for a specified amount of time and repeat those readings to distract security observation while an attack occurs \cite{mo2009secure}.
     \item \textit{Man-in-the-Middle (MitM)}: Attacker positions itself between two nodes to either eavesdrop or or impersonate one of the nodes and creates a false impression of normal system operation \cite{wlazlo2021man}. 
    \end{itemize}
    \item Confidentiality Attacks:
    \begin{itemize}
        \item \textit{Eavesdropping}: Confidential information stolen from eavesdropping from communications between sensors and controllers \cite{duo2022survey}.
    \end{itemize}
\end{itemize}

Attacks can be devastating ranging from operational disruptions to severe safety hazards. For example, DoS attacks can incapacitate communication networks and cause essential services to become inaccessible \cite{suprabhath2023cyber}. Koduru, et al. highlight how DoS attacks can impact economic losses and cause widespread outages, claiming that cyber attacks on power grids are ``the most dangerous and impactful phenomenon'' \cite{suprabhath2023cyber}. Integrity attacks like false data injection can corrupt data and commands within a CPS, resulting in erroneous system behavior. Several real-life cyber-attacks are feature in the paper, we refer the reader to \cite{suprabhath2023cyber} for an in-depth analysis and case-studies of impacts on CPSs from a microgrid stand point. When false data injections are introduced to a system like smart grids and specifically operations governing traffic, consequences like service disruption and financial looses can occur \cite{humayed2017cyber}. 

MitM attacks can be used to flood the radio communication of Unmanned Aerial Vehicles (UAV)~\cite{dahiya2020unmanned}. UAVs are used for applications including surveillance~\cite{kim2012cyber}, weather monitoring~\cite{gupta2013review}, unmanned attacks \cite{gudla2018defense}, disaster relief~\cite{debusk2010unmanned}, rescue operations~\cite{waharte2010supporting}, and many more. In the case of disaster relief, a MitM attack can intercept and alter the communication between the UAV and control station. Successfully interfering with this communication can prevent or delay the UAV from its current mission. In time sensitive missions, like disaster relief, this could delay critical aid and endanger lives~\cite{estrada2019uses}. MitM attacks can result in data breeches leading to sensitive data accessible to transnational criminal organizations or nations with adversarial relationships~\cite{li2020lightweight}. 

IoT is particularly susceptible to spoofing attacks. For example, robotic systems relying on a network for control can be vulnerable to various attacks that can gain access to wired/wireless communications like spoofing~\cite{yaacoub2022robotics}. In the case of FDI attacks on IoT, an attacker can exploit minor error margins tolerated by system algorithms to incrementally increase the impact of the injection which can ultimately allow the escalation of the false data to go unnoticed~\cite{bostami2019false}. Bostami et al. goes on to mention that if sensors are compromised in a home automation system, it can produce incorrect reports due to the introduction of false data~\cite{bostami2019false}. 




\subsection{Anomaly Detection}

Anomaly detection refers to the task of identifying irregular behavior, also known as anomalies or outliers, based on understanding regular behavior~\cite{Chandola:2009:Anomaly:Detection:Survey}. A widely accepted definition across applications define an anomaly as an ``observation that differs so much from other observations as to arouse suspicions that it was generated by a different mechanism''~\cite{Hawkins:1980:Identification:Outliers}.

The deployment of anomaly detection algorithms often prevent and deter the occurrence of critical events or even undesired critical conditions such as the case of cyberattacks~\cite{Moriano:2021:Burstiness:BGP, Moriano:2022:AHC}. Other anomaly detection applications range from different disciplines including network intrusions~\cite{Garcia:2009:Anomaly:Network:Intrusion:Detection}, credit card fraud~\cite{Zhang:2022:Optimized:Anomaly:Detection:Credit:Card}, tax evasion~\cite{Bolton:2001:Unsupervised:Profiling:Fraud:Detection}, route hijacking~\cite{Shi:2012:Argus}, and malware detection~\cite{Invernizzi:2012:Evilseed}. 

Anomaly detection techniques start by characterizing the regular behavior of a system against which unusual patterns are compared. Characterizing regular behavior is then a prerequisite for detecting anomalies. Based on the extent to which data labels are available for detection purposes, anomaly detection is usually performed in three different ways, including supervised, semi-supervised, and unsupervised anomaly detection. 

Supervised anomaly detection assumes labeled data for both normal and anomaly classes at the training stage. They focused then on building predictive models for distinguishing between both classes. Supervised anomaly detection presents two main challenges related to labeled data scarcity, namely data imbalanced and labeled data augmentation. The former is related to a significantly lower amount of anomalous data versus normal data. This issue has been mitigated in the machine learning literature based on data mining techniques~\cite{Weiss:1998:Learning:Predict:Rare:Events:Sequences, Joshi:2001:Classifying:Rare:Classes:Two:Phase, Vilalta:2002:Predicting:Rare:Events:Temporal:Domains, Joshi:2002:Predicting:Rare:Classes:Boosting, Chawla:2004:Learning:Imbalanced:Datasets, Phua:2004:Minority:Report:Fraud:Detection}. The former entails generating a representative proportion of data samples with anomaly labels. For this, some techniques focused on generating synthetic anomalies that are subsequently integrated with the training data~\cite{Theiler:2003:Resampling:Approach:Anomaly:Detection:Multispectral:Images, Abe:2006:Outlier:Detection:Active:Learning, Steinwart:2005:Classification:Framework:Anomaly:Detection}. 

Semi-supervised anomaly detection assumes there is only available normal labeled data. By relaxing the assumption about required anomaly labeled data they are usually more applicable than supervised techniques. For example, in automotive systems fault detection~\cite{Theissler:2017:Detecting:Known:Unknown:Faults:Automotive:Systems}, an anomaly could represent a critical fault in a vehicle leading to an accident. The usual way to go about using these methods is to build models using the available normal labeled data and then use it to detect anomalies  in the test data. Although there are also models that use exclusively anomaly data for training purposes, it reinforces the challenge of requiring a variety of anomaly data for proper detection of a variety of anomalies~\cite{Warrender:1999:Detecting:Intrusion:System:Calls, Dasgupta:2000:Comparison:Negative:Positive:Selection, Dasgupta:2002:Anomaly:Detection:Negative:Sampling}. 

Unsupervised anomaly detection do not require training data making it the most popular second line of defense for detecting anomalies. The implicit assumption with these techniques is that the amount of normal labeled data is far more abundant than anomaly labeled data is the test data for reducing false positives. As the most popular alternative for performing anomaly detection, it has been adapted for detecting anomalies in a variety of applications~\cite{Goldstein:2016:Comparative:Evaluation:Unsupervised:Anomaly:Detection, Zong:2018:Deep:Autoencoding:Uncupervised:Anomaly:Detection, Hanselmann:2020:CANet, Shahriar:2023:CANshield}. 

\subsection{Concept Drift}

Concept drift refers to an scenario where the underlying relationship between the input data $X$ and the output variable $Y$ changes over time~\cite{Gama:2014:Survey:Concept:Drift}. In particular, let a training sample at time $t$ be represented by the couple $\{\bm{x}_{t}, y_{t}\}$, which is generated by hidden joint distribution $P_{t} (X, Y)$. Using Bayes theorem, $P_{t} (X, Y)$ can be expressed as $P_{t} (X, Y) = P_{t} (Y | X) P_{t} (X)$, where the expressions $P_{t} (Y | X)$ and $P_{t} (X)$ represent the posterior probability of a label given the sample and the prior distribution of the data samples. Here, the joint probability $P_{t}(X, Y)$ is known as the \emph{concept}. Therefore, a concept drift happens at time $t+1$ if $P_{t} (X, Y) \neq P_{t+1} (X, Y)$.

Concept drift is very common when dealing with data streams where a data source produces data samples sequentially, in real-time, and potentially at very high speed~\cite{loeffel2017adaptive}. Concept drifts have been characterized statistically as real and virtual. On the one hand, a real concept drift happens when the posterior distribution of the labels change or $P_{t}(Y | X) \neq P_{t+1} (Y | X)$ when $P_{t} (X) = P_{t+1} (X)$. This means that the probability of a label associated with a sample has changed. On the other hand, a virtual concept drift happens when the prior distribution of the samples change or $P_{t}(Y | X) = P_{t+1} (Y | X)$ when $P_{t} (X) \neq P_{t+1} (X)$. This means that the probability of occurrence of data samples has changed. 

\subsection{Learning}

Learning algorithms are usually classified as offline and online. Offline algorithms refers to the case when the whole dataset of samples on which the algorithm is learning is available at training time. Thus, only when training is completed the model can be used for prediction. On the other hand, online algorithms learn by by processing data samples one by one without assuming to have the whole training dataset available at the beginning. In the context of adaptive anomaly detection, we focus exclusively in online algorithms.

Here learning refers to the mechanisms that are used to learn from new data samples and update the prediction models. We discuss three different aspects related to learning: (1) learning mode, (2) adaptation methods, and (3) ensemble methods based on the work by Gama et al.~\cite{Gama:2014:Survey:Concept:Drift}. 

Learning mode refers to the process of incorporating the knowledge of new data samples into the prediction models. Here, two main options can be operationalized: retraining and incremental. On the one hand, retraining assumes the existence of enough memory for data storage. Specifically, after training a model with the available data, whenever a new data sample arrives, the previous model is discarded. New data samples are then merged with previous data so a new model can be learned on this expanded dataset~\cite{Wu:2020:Deltagrad:Rapid:Retraining:ML, Mahadevan:2024:Cost:Aware:Retraining:ML}. On the other hand, incremental learning focuses on updating the model based on the most recent data. In doing so, algorithms process the input one-by-one and update models accordingly. Incremental learning is also used to retain knowledge acquired from previous classes~\cite{Wu:2019:Large:Scale:Incremental:Learning, Van:2022:Three:Types:Incremental:Learning}. 

Adaptation refers to control strategies to perform the action of adapting prediction models. Adaptation methods are usually classified as blind and informed. On the one hand, in the blind category, algorithms simply adapt models without any explicit signal suggesting to do so. Common to blind adaptation is the use of a fixed sliding window of size $\omega$ that frequently update the model by training on the last $\omega$ samples~\cite{Klinkenberg:1998:Adaptive:Information:Filtering:Learning:Concept:Drifts, Lanquillon:2001:Enhancing:Classification:Information:Filtering}. As the blind strategy is proactive, they tend to be slow when reacting to concept drifts. This is because they will forget about old concepts independent if they are happening or not. On the other hand, in the informed category, strategies are reactive depending on a signal suggesting to update the models~\cite{Bifet:2006:Kalman:Filter:Learning:Data:Streams}. Note that triggers can be based on detector's output independent of the adaptation strategy but they could also be integrated with the adaptation strategy~\cite{Gama:2006:Decision:Data:Mining:Streams}; in addition, triggers can be based on specific data descriptors~\cite{Widmer:1996:Learning:Concept:Drift:Hidden:Context}.   

Ensembles combine the knowledge of multiple learners to predict outcomes. The main advantage of ensembles is the idea of creating a diversified base of knowledge using the collective wisdom of multiple learners so that learners' weaknesses can be mitigated. Predictions are then obtained by aggregating individual learners' predictions. Note that the final prediction can be obtained using different aggregation methods. Popular approaches include the idea of having learners to vote under the assumption of same vote weight. Here the prediction of the ensemble is given by the majority vote. Another popular option is is to weight votes based recent learners performance. Note that in this case, weights will evolve over time ensuring that top learner performers are assigned a higher weight. 

Note that in the case of steaming data the categorization of ensembles vary. Following Kuncheva's work~\cite{Kuncheva:2004:Classifier:Ensemble:Changing:Environments}, ensemble classification include (1) modifying the combination rule of ensembles as concept drift happens (or dynamic combination)~\cite{Tsymbal:2006:Handling:Local:Concept:Drift}; (2) allowing base learners to update their models using latest processed data (or continuous update)~\cite{Fern:2003:Online:Ensemble:Learning, Rodriguez:2008:Combining:Classification:Changing:Environments}; (3) removing or adding learners to the ensemble (or structural update)~\cite{Street:2001:Streaming:Ensemble:Algorithm, Bouchachia:2011:Incremental:Learning:Multi:Adaptation}.

\subsection{Data Management}

Data management is a critical component of modern CPS; the topics covered here pertaining to data management are collection of the data, processing of said data, and retention of data. CPS collect data from sources like sensors in real-time creating what is called a data stream. A data stream is a continuous (unbound) flow of data that generates observations in sequence and in real-time without a predefined end~\cite{loeffel2017adaptive, golab2003issues, bahri2021data}. From \cite{loeffel2017adaptive}, we formalize data stream as the following. At time $t_0$, a data stream will begin to produce unlabeled observations $\{x_{t0}, x_{t1}, x_{t2}, ...\}$, delivered sequentially at consistent time intervals ($\forall k \in \mathbb{N} : t_{k+1}-t_k=u$ where $u \in \mathbb{R}^+$ is a constant) and continues indefinitely. Observation $x_t$ is corresponded by label $y_t$, Loeffel mentions that $y_t$ is assumed to always be available before the subsequent observation $x_{t+1}$, which we adopt as well. In the realm of machine learning, the objective remains consistent with traditional offline methodologies. The main task is to develop a function $h:X \to Y$ that can effectively forecast the label $y_t$ that corresponds to each observational input $x_t$~\cite{loeffel2017adaptive}. 

As hardware and software advances, data streams are generated from an ever increasing variety of sources. Some popular sources of data streams are IoT devices like sensor networks~\cite{golab2003issues}, a wide variety of embedded systems~\cite{muller2010data} like ECUs~\cite{verma2024comprehensive}, computer network traffic like data packets and logs~\cite{gaber2005mining}. Data stream management systems must handle the unique characteristics of data streams and continuous queries by supporting order-based and time-based operations, such as queries over moving windows~\cite{golab2003issues}. Due to the inability to store complete streams, these systems often use approximate summary structures (synopses or digests) which may not provide exact answers, and they avoid blocking operators to ensure timely results~\cite{golab2003issues}. We can see some of these requirement present in real-world applications like wireless sensor networks where numerous nodes with limited computing and communication capabilities collaborate to perform tasks such as data acquisition or event detection~\cite{muller2010data}.  

Bahri et al.~\cite{bahri2021data} state that data streams mining faces several common challenges across tasks, possibly the most important is dealing with ever evolving data streams that deliver instances quickly and require scalable algorithms capable of processing data dynamically. Loeffel et al.~\cite{loeffel2017adaptive} claim that online learning algorithms are essential for handling sequential observations and allow models to continuously update with new data points without access to the entire dataset. These algorithms must also efficiently manage constrained memory due to the impractically of storing all past observations for potentially infinite data streams. Gama et al.~\cite{Gama:2014:Survey:Concept:Drift} state that a major challenge in data streams is concept drift where changes in data distribution require adaptive algorithms to maintain accuracy and relevance. In this context, memory management involves updating predictive models with new information while forgetting outdated data. Short-term memory deals with which data to learn from and how to discard old data which utilizes the assumption that recent data is most informative. To handle the demanding processing requirements of modern data streams, innovative processing techniques have been employed throughout literature. Gama et al. also mention that online learning algorithms typically use single-example memory and processes new observations in sequence while updating the model as needed. These methods typically lack explicit forgetting mechanics and may adapt slowly to abrupt changes. Multiple-example memory approaches (sliding windows, etc.) maintain a set of recent examples to build models that balance fast adaptation with stability through fixed or variable window lengths. 

A crucial element in data streams is the forgetting mechanism used for processing evolving data and discarding outdated observations for model relevance. Gama et al.~\cite{Gama:2014:Survey:Concept:Drift} outline two main forgetting mechanisms, abrupt and gradual. Abrupt forgetting uses sliding windows to keep only recent data and gradual forgetting retains all data but decreases the importance of older observations using a weighting system. Gradual forgetting uses an aggregation function $G(X,S)$ and newer statistics are computed as $S=G(X_i,\alpha_{i-1})$, where $\alpha \in (0,1)$~\cite{Gama:2014:Survey:Concept:Drift} is a fading factor.

\subsection{Evaluation Metrics}

To evaluate performance of learning algorithms within the data stream environment, appropriate performance evaluations and metrics must be selected in coordination with the learning task's objective. It is particularly important to determine which observations are used for training and testing~\cite{Gama:2014:Survey:Concept:Drift}. Loeffel~\cite{loeffel2017adaptive} mentions that traditional cross-validation isn't suitable for evolving data streams due to temporal order dependencies. Two common methods are used to evaluate time-ordered data; Holdout and Prequential~\cite{loeffel2017adaptive, gama2009issues, Gama:2014:Survey:Concept:Drift, cerqueira2020evaluating}. Holdout method works by training a model on a set of observations and testing on a subsequent observations at regular intervals~\cite{loeffel2017adaptive}. Here, the algorithm is trained with a set of observations $\{t_1, \ldots, t_k\}$ which produces a model $h_{t_k}$. This is then tested on a subsequent set $\{t_{k+1}, \ldots, t_{k+l}\}$~\cite{loeffel2017adaptive}. Next, the model is updated with the next set of observations $\{t_{k+1}, \ldots, t_2k\}$ which results in a new model $h_{t_{2k}}$ and is tested again on $\{t_{2k+1}, \ldots, t_{2k+l}\}$. This entire procedure repeats until all observations are used~\cite{loeffel2017adaptive}.

Prequential evaluation tests new observations before using its label for training (updating the model) which allows for continuous performance assessment~\cite{loeffel2017adaptive}.  Three prequential evaluation methods are typically used; landmark window (Interleaved Test-Then-Train), sliding windows, and forgetting mechanisms~\cite{loeffel2017adaptive}. The landmark window error is computed as the average of all prediction errors at time $T$ since the beginning of the streaming data given as $L_T=\frac{1}{T}\sum_{t=1}^n\ell(y_t, \hat{y}_t)$~\cite{loeffel2017adaptive}. The fading factor technique applies a decay factor, $\alpha$, to previous errors. This technique offers tracking of recent performance changes using the formula $L_T=\frac{S_T}{N_T}$ where, $S_T=\ell_t +\alpha S_{T-1}$ and $N_T=1 +\alpha N_{T-1}$~\cite{loeffel2017adaptive}. Finally, sliding windows evaluate performance over the latest $k$ observations expressed as $L_T=\frac{1}{k}\sum_{t=T-k+1}^T \ell_t$~\cite{loeffel2017adaptive}. This highlights algorithm reactivity to concept changes similar to fading factor~\cite{loeffel2017adaptive}. 

Traditional metrics are used like precision, recall, and sensitivity for accuracy measures with metrics like root mean square error (RMSE) and mean absolute scaled errors for regression objectives.~\cite{Gama:2014:Survey:Concept:Drift}. Baseline approaches like moving average used during prediction of time-series are especially important for comparing the effectiveness of intelligent adaptive techniques against simple predictions. Streaming settings to consider are the computational cost (RAM-hours) and class imbalance (kappa statistic). Gama et al.~\cite{Gama:2014:Survey:Concept:Drift} mention that change detection performance is evaluated using measures like the probability of true change detection, probability of false alarms, and detection delay. Utilizing these metrics during inference can help researchers or practitioners assess the impact of change detection on the overall performance of adaptive models especially with known or synthetic drifts. 

When evaluating adaptive learning techniques across multiple aligned data streams, it is essential to apply and estimate the performance of these models on individual but related data streams using methods like cross-validation~\cite{Gama:2014:Survey:Concept:Drift}. This approach is particularly useful for models applied across similar feature spaces and allows for parameter tuning and validation across subsets of streams. Statistical significance is important when comparing classifiers with many researchers and practitioners using the McNemar test to assess differences in error rates between two classifiers. The McNemar statistic $M$ is given as: $M=sign(a-b) \times \frac{(a-b)^2}{a+b}$, where $a$ is the number of instances misclassified by the first classifier and correctly classified by the second, and $b$ represents the number of instances where the second classifier misclassifies data that the first classifier correctly classifies.~\cite{Gama:2014:Survey:Concept:Drift}. The McNemar test follows the $\chi^2$ distribution. When more than two classifiers are used, Gama et al. point to the Nemenyi test that evaluates performance differences using average ranks across datasets which determines significant differences with a critical difference (CD) formula: $CD=q_\alpha \sqrt{\frac{k(k+1)}{6N}}$, where the number of learners is $k$, the number of datasets is $N$, and $q_\alpha$ are the critical values. 
\section{Related Work} \label{sec:related work}



As the field of AAD in CPS has matured over the last decade, several survey papers have been published with both broad and narrow scope in this field. The works discussed in this section were identified and compiled during the comprehensive literature search detailed in Section \ref{sec:Methods}.

Saad et al.~\cite{Saad:2019:Review:Mitigation:Attacks:SmartGrids} analyzed cybersecurity challenges in smart grids. They examined vulnerabilities, security needs, detection techniques, countermeasures, and secure communication protocols. Their study emphasized integrating advanced communication and computing technologies to improve reliability and efficiency. They proposed solutions to mitigate cyberattacks, enhancing the security of power networks. Meng et al.~\cite{Zhang:2019:Review:FDI:Against:SG:Estimation} summarized recent advances in false data injection attacks (FDIA) targeting smart grid state estimation. They reviewed FDIA construction methods, detection strategies, and defense mechanisms. They also outlined future directions, such as applying FDIAs to alternating current (AC) state estimation and using data-driven models for FDIA detection. Syrmakesis et al.~\cite{Syrmakesis:2022:Classifying:Resilience:Approaches:Smart:Grid} classified cyber-resilience methods for protecting smart grids. They reviewed approaches addressing cyberattacks and anomalies. Their taxonomy supports research on cyber-resilience and identifies promising directions for smart grid security.

Rojas and Rauch~\cite{Rojas:2019:Review:Enablers:Smart:Manufacturing} conducted a systematic review of 165 papers on CPS production trends. They highlighted the importance of connectivity and control systems in manufacturing. Their work grouped papers into six categories, including cybersecurity enablers for smart manufacturing. Zeadally et al.~\cite{Zeadally:2019:Self:Adaptation:CPS} reviewed 12 self-adaptive approaches for managing large-scale CPS. They discussed the strengths, weaknesses, and implementation techniques for self-adaptive mechanisms across CPS architectural layers, including physical, network, and cyber. Rosenberg et al.~\cite{Rosenberg:2021:Adversarial:ML:Attacks:Defense:Cyber} reviewed 58 papers on adversarial attacks and defenses in cybersecurity. They proposed a taxonomy based on attack stages, goals, and capabilities. Their work highlighted future research needs in adversarial machine learning (ML). Jamal et al.~\cite{Jamal:2023:Review:Security:Analysis:CPS:ML} reviewed ML and deep learning (DL) techniques for CPS security. They identified challenges, reviewed existing methods, and proposed future directions for artificial intelligence (AI)-based CPS protection against cyber threats. Pekaric et al.~\cite{Pekaric:2023:Systematic:Security:Safety:Self:Adaptive:Systems} reviewed 21 papers on safety and security in self-adaptive systems. They found that current approaches rarely model both aspects together and often rely on simulations with simplified use cases. Koay et al.~\cite{Koay:2023:ML:ICS} listed vulnerabilities and cyberattacks in ICS. They compared ML methods for attack detection and highlighted challenges like limited datasets and risks of adversarial attacks. They proposed urgent research directions.

Mahapatra et al.~\cite{Mahapatra:2020:Survey:IoT:Taxonomy:Challenges} created a taxonomy for secure IoT communication. They classified architectures, communication topologies, and secure transmission methods, including cluster-based and blockchain-based techniques. They identified research challenges and opportunities to enhance IoT security. Stoyanova et al.~\cite{Stoyanova:2020:Survey:IoT:Forensics} reviewed IoT forensics and its challenges, such as device diversity, encryption, and cloud reliance. They examined theoretical models and proposed frameworks using blockchain for evidence integrity. They emphasized the need for standardized forensic processes and readiness strategies. Dai and Boroomand~\cite{Dai:2022:Review:AI:Security:Big:Data} reviewed 58 papers on security issues in big data systems. They focused on AI techniques like DL and multi-agent systems for detecting and mitigating attacks. They mapped these methods to security strategies and evaluation models. Huang et al.~\cite{Huang:2022:RL:Feedback:Enabled:Cyber} reviewed reinforcement learning (RL) for cyber-resilience. They discussed vulnerabilities, including posture- and information-related issues, and applications like moving target defense. They also analyzed vulnerabilities in RL systems under adversarial attacks. Alaghbari et al.~\cite{Alaghbari:2022:Complex:Event:Processing:Datacenter} surveyed complex event processing (CEP) in CPS security. They highlighted how combining CEP with ML enhances intrusion detection. They also discussed open issues in CEP applications for cybersecurity. Lin et al.~\cite{Lin:2023:Security:5G:IoT:Factories} surveyed security and privacy issues in 5G-industrial IoT (IIoT) factories. They reviewed approaches using DL, RL, and blockchain. Their work identified research opportunities for securing 5G-enabled industrial systems.

Grim et al.~\cite{Grimm:2021:Context:Aware:Vehicles:Fleet} reviewed 50 papers on adaptive and intelligent security for vehicles and fleets. They identified open research areas and developed a taxonomy of contextual information categories. Their work aims to guide future developments in automotive security. Strandberg et al.~\cite{Strandberg:2022:Systematic:Review:Automotive:Digital:Forensics} conducted a systematic review of 67 papers on automotive digital forensics. They categorized papers into technical solutions and surveys and mapped forensic data to security properties and stakeholders. Their findings are relevant to CPS like smart cities.

Cooper et al.~\cite{Cooper:2023:Anomaly:Detection:Power:System:State:Estimation} reviewed anomaly detection methods for power system state estimation. They connected traditional data-driven methods with modern approaches addressing new cyber threats. They proposed directions for future research, including dynamic load profiles and asynchronous measurements.

Table~\ref{table: related work} lists and identifies prior surveys related to AAD in CPS. It also outlines publication details and the contribution of existing surveys in the area. While there have been several valuable works reviewing different aspects of AAD in CPS, including adaptability~\cite{Zeadally:2019:Self:Adaptation:CPS, Pekaric:2023:Systematic:Security:Safety:Self:Adaptive:Systems} and online learning~\cite{Rosenberg:2021:Adversarial:ML:Attacks:Defense:Cyber, Huang:2022:RL:Feedback:Enabled:Cyber}, they do not provide an extensive analysis of other critical aspects related to these tasks including data management and concept drift. For example, Zeadally et al.~\cite{Zeadally:2019:Self:Adaptation:CPS} focuses on discussing techniques that enable self-adaptation capabilities within CPS at different architectural layers. However, it lacks insight into the role of different data management strategies to handle streaming data conditions, as well as the categorization of adaptive techniques per application. To address these limitations and provide an in-depth understanding of state-of-the-art on AAD in CPS, we conduct a systematic and extensive survey of related literature. By comprehensively collecting $397$ papers and systematically analyzing $65$ papers (47 research papers and 18 surveys) from top journals and conferences, our SLR aims to provide a holistic summary of AAD methods and its broad applicability within the CPS domain. This approach enables a more nuanced analysis of the field highlighting strengths and limitations of different approaches. This contributes to identify most promising directions for future research. 
\begin{table*}[htp!]
\centering
\caption{State-of-the-art surveys related to adaptive anomaly detection in CPS.}
\label{table: related work}
\begin{adjustbox}{max width=\textwidth}
\begin{tabular}{|p{0.5cm}|p{3cm}|p{1cm}|p{2cm}|p{2.5cm}|p{4.0cm}|p{1.5cm}|}
\hline
{No.} & {Reference} & {Year} & {Time Frame} & {No. Papers Surveyed} & {Survey Topic} & {Application} \\
\hline
1 & Saad et al.~\cite{Saad:2019:Review:Mitigation:Attacks:SmartGrids} & 2019 & Not specified & Not specified & Modern strategies for  & Smart grid \\
& & & & &  mitigating cyberattacks & \\
2 & Meng et al.~\cite{Zhang:2019:Review:FDI:Against:SG:Estimation} & 2019 & Not specified & Not specified & Recent advances against FDIAs & Smart grid  \\
3 & Rojas and Rauch~\cite{Rojas:2019:Review:Enablers:Smart:Manufacturing} & 2019 & 2012-2017 & 165 & Smart manufacturing control & CPS  \\
4 & Zeadally et al.~\cite{Zeadally:2019:Self:Adaptation:CPS}	& 2019 & Not specified & 12 & Self-adaptive mechanisms & CPS  \\
5 & Mahapatra et al.~\cite{Mahapatra:2020:Survey:IoT:Taxonomy:Challenges} & 2020 & Not specified & Not specified & Secure transmission in IoT & IoT  \\
6 & Stoyanova et al.~\cite{Stoyanova:2020:Survey:IoT:Forensics} & 2020 & Not specified & Not specified & IoT forensics & IoT \\
7 & Rosenberg et al.~\cite{Rosenberg:2021:Adversarial:ML:Attacks:Defense:Cyber} & 2021 & Not specified & Not specified & Adversarial attacks and defenses & CPS \\
8 & Grim et al.~\cite{Grimm:2021:Context:Aware:Vehicles:Fleet} & 2021 & Not specified & 50 & Context-aware security & Vehicle \\
9 & Strandberg et al.~\cite{Strandberg:2022:Systematic:Review:Automotive:Digital:Forensics} & 2022 & 2006-2021 & 67 & Automotive digital forensics & Vehicle \\
10 & Dai and Boroomand~\cite{Dai:2022:Review:AI:Security:Big:Data} & 2022 & 2006-2021 & 58 & AI-driven security for big data & IoT \\
11 & Syrmakesis et al.~\cite{Syrmakesis:2022:Classifying:Resilience:Approaches:Smart:Grid} & 2022 & Not specified & Not specified & Classification of cyber resilience & Smart grid \\
& & & & &  methods & \\
12 & Huang et al.~\cite{Huang:2022:RL:Feedback:Enabled:Cyber} & 2022 & Not specified & Not specified & RL for cyber resilience & IoT \\
13 & Alaghbari et al.~\cite{Alaghbari:2022:Complex:Event:Processing:Datacenter} & 2022 & Not specified & Not specified & Complex event processing (CEP) & IoT \\
14 & Jamal et al.~\cite{Jamal:2023:Review:Security:Analysis:CPS:ML} & 2023 & Not specified & Not specified & Analysis of ML and DL applications & CPS \\
15 & Pekaric et al.~\cite{Pekaric:2023:Systematic:Security:Safety:Self:Adaptive:Systems} & 2023 & 2000-2020 & 21 & Security of self-adaptive systems  & CPS \\
16 & Cooper et al.~\cite{Cooper:2023:Anomaly:Detection:Power:System:State:Estimation} & 2023 & Not specified & Not specified & Anomaly detection for power system  & Power grid \\
& & & & &  state estimation & \\
17 & Lin et al.~\cite{Lin:2023:Security:5G:IoT:Factories} & 2023 & 2018-2021	& 22 & Security and privacy in 5G-IIoT & IoT \\
18 & Koay et al.~\cite{Koay:2023:ML:ICS} & 2023 & 2017-2022 & 30 & Attacks and defenses in ICS & CPS \\

\hline
\end{tabular}
\end{adjustbox}
\end{table*}

\section{SLR Methodology} \label{sec:Methods}

\subsection{Survey Method} 
This section discusses the scope and SLR used in this work. We adapt Okoli's \cite{okoli2015guide} review process for creating a standalone SLR and designed it with precision to ensure the comprehensive collection of relevant articles while maintaining methodological rigor. This SLR centers on summarizing the literature on AAD methods for CPS, focusing on the mechanisms and models used, rather than providing a comprehensive taxonomy of the specific threats addressed within the field (e.g., DoS, FDI, or advanced persistent threats (APT)). Here, the review process consists of several sequential steps including planning, selection, extraction, and execution that ultimately organize the review process into a reproducible SLR. 

\subsubsection{Eligibility Criteria}
To maintain the integrity of this SLR and ensure that the selected articles met the requisite standards of relevance and quality, we establish and follow a strict inclusion and exclusion criteria. To be included in this SLR, studies had to:

\begin{itemize}
    \itemindent=-13pt
    \item Be published in peer-reviewed journals or conference proceedings, ensuring that only high-quality, validated research was considered.
    \item Be published within the past 10 years (2013–Nov. 2023) to capture recent developments and advancements in AAD, reflecting the rapidly evolving nature of this field.
    \item Be written in English to maintain consistency in interpretation and reduce potential translation biases.
    \item Address topics relevant to AAD or its subdomains with a clear focus on adaptive, cognitive, or autonomous mechanisms for anomaly detection in CPS.
\end{itemize}

Articles were excluded if they:
\begin{itemize}
    \itemindent=-13pt
    \item Were non-peer-reviewed articles, such as editorials, letters, etc.
    \item Lacked a clear connection to AAD or cognitive approaches for CPS, ensuring that the review remained focused on its central research question.
\end{itemize}


\subsubsection{Data Sources}
This SLR was conducted by retrieving academic papers and conference proceedings from five prominent article databases: IEEE Xplore (\url{https://ieeexplore.ieee.org/}), ACM Digital Library (\url{https://dl.acm.org/}), Emerald Insight (\url{https://emerald.com/insight/}), Springer Link (\url{https://link.springer.com/}), and ScienceDirect (\url{https://sciencedirect.com/search}). These databases were selected due to their extensive coverage of fields related to cybersecurity, adaptive systems, and AI, ensuring a comprehensive exploration of relevant literature. Each database was searched for publications covering the period from 2013 to November 2023 to align with the eligibility criteria. In addition to electronic searches within these databases, supplementary strategies (forward/backward search~\cite{wohlin2014guidelines}) were applied to identify additional key studies. This approach enabled us to capture influential works not readily indexed by standard database searches and enhanced the comprehensiveness of our SLR.

\subsubsection{Search Strategy}
\begin{figure*}[htp]
    \centering
    \includegraphics[width=\textwidth]{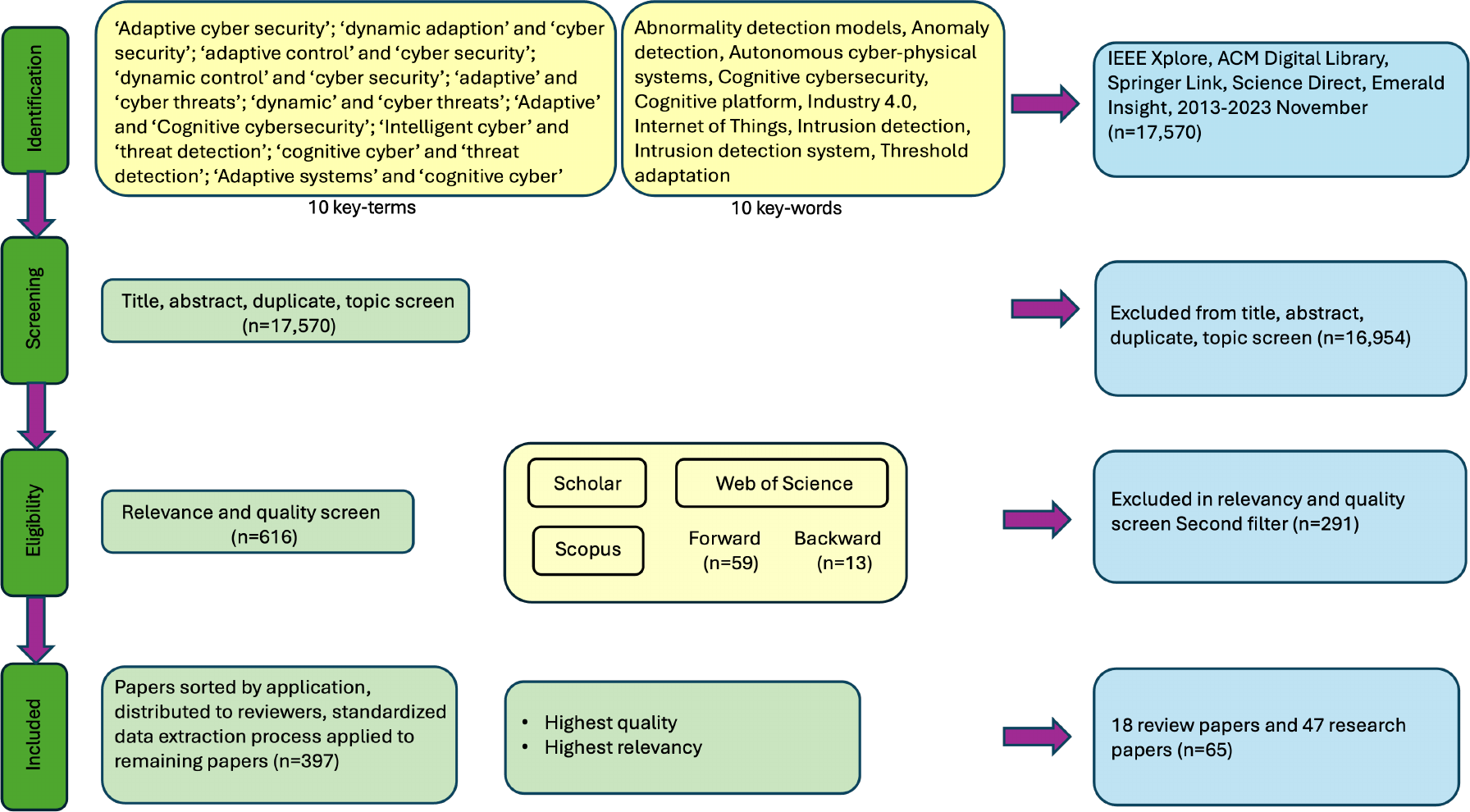}
    \caption{Flow chart of the search and selection process, highlighting stages from initial identification through screening, eligibility, and final inclusion. This sequential filtering ensures a rigorous selection of high-quality papers. Here, yellow bubbles highlight a process that adds papers to the selection pool, green bubbles highlight a process that screens papers, and blue bubbles highlight the result.}
    \label{fig:des-dia}
\end{figure*}

Figure \ref{fig:des-dia} highlights the initial search process, which first establishes 10 key-terms including: \texttt{adaptive cyber security}; \texttt{dynamic adaption AND cyber security}; \texttt{adaptive control AND cyber security}; \texttt{dynamic control AND cyber security}; \texttt{adaptive AND cyber threats}; \texttt{dynamic AND cyber threats}; \texttt{adaptive AND cognitive cybersecurity}; \texttt{intelligent cyber AND threat detection}; \texttt{cognitive cyber AND threat detection}; \texttt{Adaptive systems AND cognitive cyber}. 10 keywords were searched as well, including: \texttt{abnormality detection models}, \texttt{anomaly detection}, \texttt{autonomous cyber-physical systems}, \texttt{cognitive cybersecurity}, \texttt{cognitive platform}, \texttt{industry 4.0}, \texttt{internet of things}, \texttt{intrusion detection}, \texttt{intrusion detection system}, \texttt{threshold adaptation}. Adjustments were made to improve search efficacy based on each database’s functionality. For example, the search term ``adaptive security'' was used in the ACM Digital Library to retrieve a broader set of results, while ScienceDirect required expanded criteria to title/abstract/author-specified keywords for higher relevancy.

To enhance the comprehensiveness of the SLR, we performed a forward and backward search on a set of selected high-impact and high-quality papers. The set of core papers were used in congruence with a research analytics tool called SciVal from Scopus, which provided us with accurate and up-to-date citation numbers for each paper. After setting a threshold at the 99th percentile for citation numbers, we identified four papers that met or exceeded this criterion. These core papers became the cornerstone of our forward and backward search, allowing us to trace their references (backward search) and discover the subsequent papers that cited them (forward search). Utilizing this approach was instrumental in ensuring that our SLR encompassed foundational works and recent advancements within the field of AAD. 

\subsubsection{Study Selection}
\begin{figure*}[htp]
    \centering
    \includegraphics[width=\textwidth]{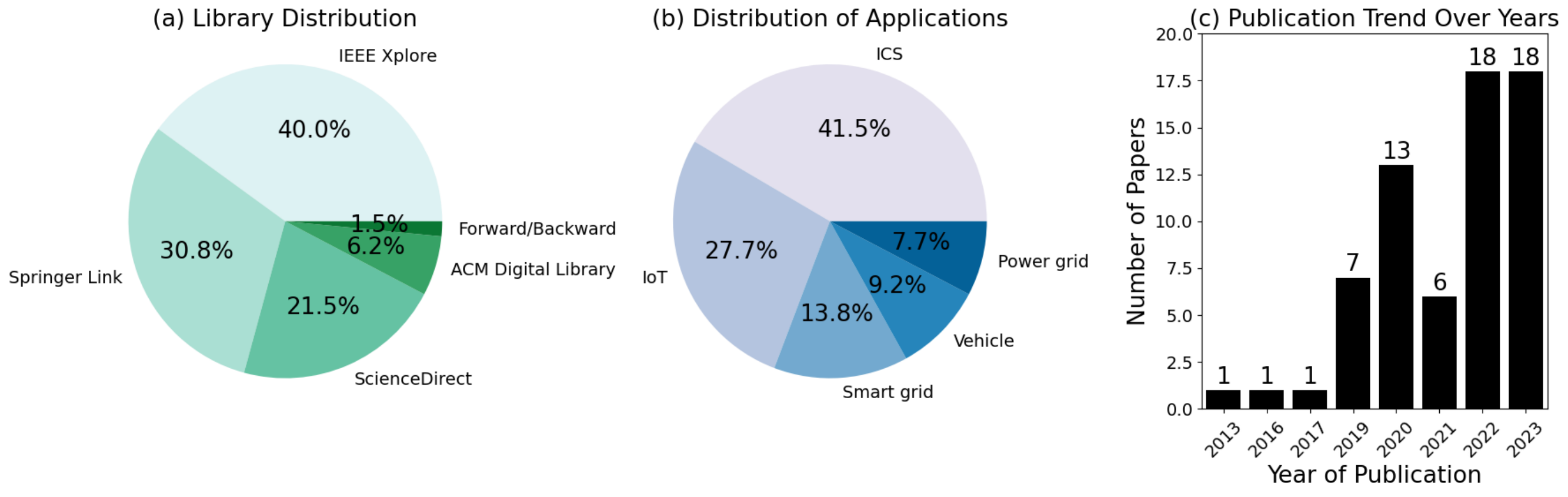}
    \caption{Distributions across selected papers in the review. (a) Digital libraries distribution, with IEEE Xplore, Springer Link, Science Direct, and ACM Digital Libraries as primary sources. (b) Distribution of applications including CPS, IoT, smart grid, vehicle, and energy (c) Annual distribution of reviewed papers from 2013 to 2023.}
    \label{fig:trends}
\end{figure*}

From here, we create and manually apply a tagging system. The purpose of this step was to be able to systematical filter relevant papers based on what a papers' main focus was. Figure \ref{fig:trends} highlights the distribution of applications and digital library used amongst the papers based on our tagging system. After filtering for relevance and removing duplicates is performed with the first screening. The tagging system enabled us to organize papers systematically, supporting the identification of commonalities and trends in the literature. Pie charts were used to represent distributions across different digital libraries and application areas, facilitating a concise visualization of contributions in each category.

\subsubsection{Data Collection Process}
The final process of this methodology involved the extraction of all the papers selected for review. To approach this process from a systematic stand point, we constructed a data extraction form for reviewers to extract relevant information from each paper and normalize the process to reduce individual reviewer bias. The extraction process was inspired by Gama et al.~\cite{loeffel2017adaptive}, which lays out the basic ingredients for AAD. The data collection process involved the creation of a standardized data extraction template. This template was structured into several ``blocks" to capture key information from each paper, including the general information, dataset, model, data management, learning, and paper quality blocks. Each block contained specific fields, some with selectable options (e.g., ``application tags'') and others allowing for manual input to capture nuanced information. For example, within the ``learning block'' are specific items for the authors to address including; learning paradigm, learning mode, model adaptation, and ensemble. Each of these items within this block aim to extract specific information from each paper. The final block determines the overall quality of the paper for establishing the potential usage of a paper as a featured paper for final review. This standardized approach minimized individual reviewer bias and ensured consistent data extraction. To ensure only high-quality studies were included, we established a set of threshold criteria. Each paper was evaluated based on its relevance to AAD, credibility of findings, and specific contributions. This rigorous filtering process led to the final selection of 65 high-quality and directly related papers, of which 18 were review papers. 

\subsection{Trend Analysis}
An important aspect of this SLR is to examine trends in the selected literature, allowing the data to highlight the primary focus areas and the evolution of AAD for CPS research. This analysis provides a structured view of the research landscape and informs the thematic focus of this SLR. Figure \ref{fig:trends}(a) shows the library distribution revealing that the majority of selected papers were sourced from IEEE Xplore (40.0\%), followed by Springer Link (30.8\%) and Science Direct (21.5\%). This distribution suggests that these databases are central repositories for cybersecurity research, particularly for studies focused on AAD aspects. The prominence of IEEE highlights its significant role in the dissemination of research within technical and engineering domains and aligns well with the focus on adaptive cybersecurity solutions.

Figure \ref{fig:trends}(b) displays the application distribution across the selected papers, with ICS emerging as the dominant application area, constituting 41.5\% of the studies. This finding underscores the importance of ICS in AAD since ICS environments often involve complex interactions between physical and digital components. IoT (27.7\%) and smart grid (13.8\%) also represent significant portions of the application distribution, reflecting an increase in cybersecurity challenges within interconnected and critical infrastructure systems. Remaining applications like vehicles (9.2\%) and the power grid (7.7\%), further demonstrate the multidisciplinary nature of AAD research. This demonstrates vulnerabilities across various sectors where adaptive methods are essential to managing dynamic threats.

Figure \ref{fig:trends}(c) shows the publication trend over the years. This indicates a noticeable increase in research activity from 2019 onwards and highlights peaks in 2022 and 2023 (each year contributing 18 papers to the final selection). This upward trend reflects a growing recognition of the need for adaptive and cognitive approaches to cybersecurity, especially as cyber threats become more sophisticated and pervasive. The surge in publications over the past five years aligns with advancements in AI and ML, which have enabled the development of more complex and responsive cybersecurity solutions.

\section{AAD in CPS} \label{sec:Results}

Anomaly detection methods can be categorized into two categories as offline and online anomaly detection based the nature of data being processed~\cite{Chandola:2009:Anomaly:Detection:Survey, Odiathevar:2019:Hybrid:Offline:Online:IDS}. On the one hand, offline methods are trained from a static dataset. Most offline methods focus on thresholding requiring extensive training~\cite{Ibidunmoye:2017:AAD:Performance:Metric:Streams}. Although offline methods are able to identify complex patterns, they need to be retained whenever model's performance deteriorates. Therefore, as data streams imply rapid contextual changes from a learnt baseline, offline methods are usually prone to produce a significant number of false positives. On the other hand, online methods are usually based on incrementally learning from time windows as new data arrives~\cite{Moriano:2024:Benchmarking:Unsupervised:Online:IDS}. This implies that online methods are better suited to tackle concept drift. AAD based on offline and online methods have been successfully used by researchers to detect cyberattacks in CPS~\cite{Wang:2020:Multi:Agent:Attack:Resilient:Smart:Grid, Gyamfi:2022:Novel:Online:IDS:IoT:OI-SVDD:AS-ELM, Alsulami:2023:Security:Autonomous:Vehicle:Transfer:Learning}.

\begin{figure*}[!htp]
\centering
\includegraphics[width=0.90\textwidth]{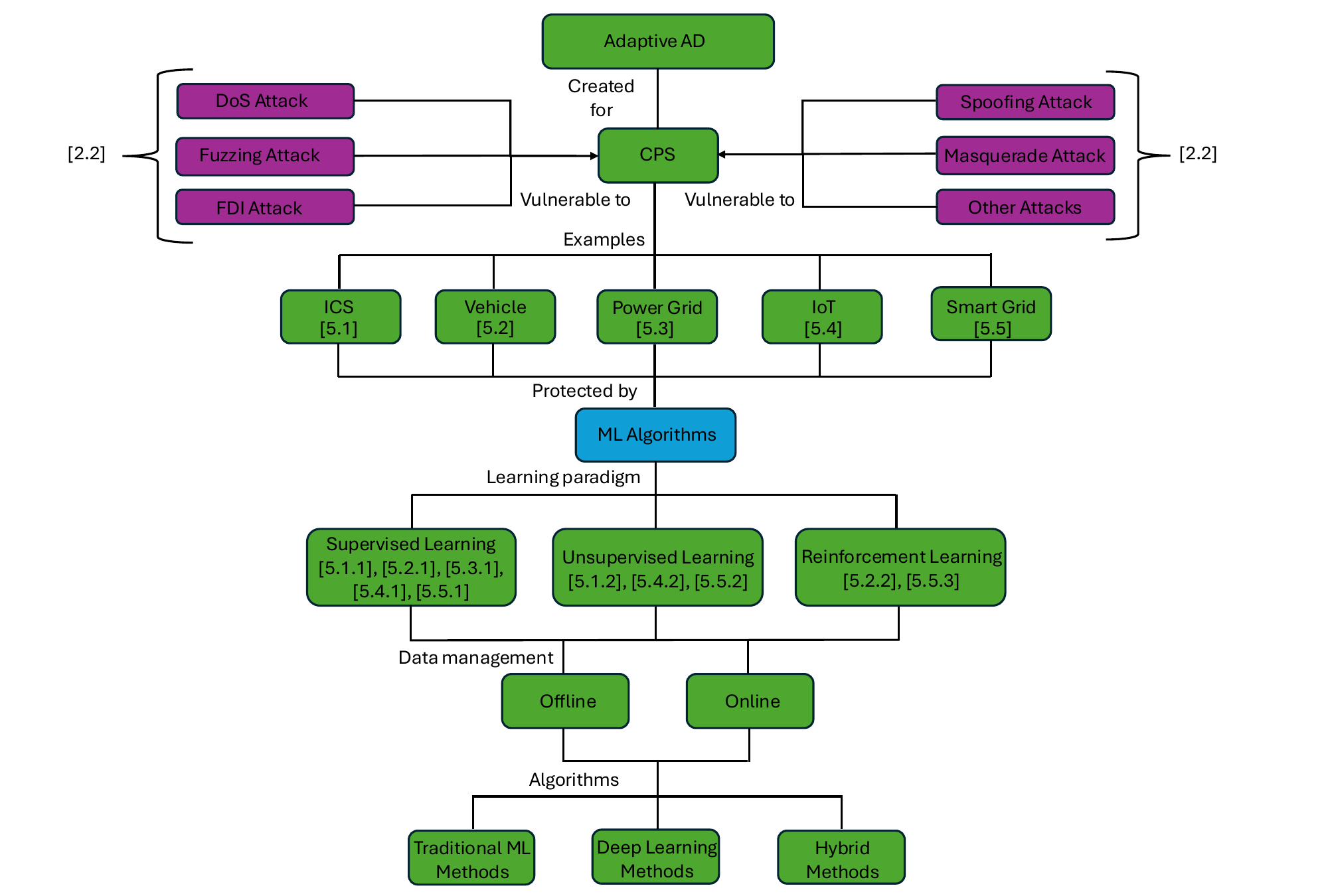}
\caption{AAD for CPS taxonomy. The numbers indicate the sections that cover each topic in the taxonomy.}
\label{fig:slr taxonomy}
\end{figure*}

The taxonomy of papers reviewed in this SLR review is shown in Figure~\ref{fig:slr taxonomy}. CPS are vulnerable to cyberattacks such as denial of service (DoS), fuzzing, FDIA, spoofing, and masquerade attacks, among others. In the literature, authors reported other attacks scenarios including advanced persistent threats, jamming, and ransomware. For a comprehensive review of other attacks on CPS, we refer the reader to the work by Yampolskiy et al.~\cite{Yampolskiy:2013:Taxonomy:Attacks:CPS} and Kim et al.~\cite{kim2022survey}. Based on the type of CPS, these attacks target different applications. We then consider the domain application to classify existing work. Specifically, we focus on the more prevalent CPS applications including ICS, vehicle, power grid, IoT, and smart grid~\cite{humayed2017cyber}. While surveying the literature, it was observed that studies commonly sorted applications based on certain themes. We summarized these papers by establishing distinct categories. Studies pertaining to ICS focused on centralized, real-time automation of industrial processes, prioritizing operational reliability and safety using fixed-location infrastructure. Papers focused on vehicles encompassed connected and autonomous systems, operating in dynamic, mobile environments where rapid decision-making and adaptability are critical. Power grid investigations ensured stable energy transmission and distribution with an emphasis on real-time monitoring for operational security. Papers explicitly focused on IoT systems connected heterogeneous devices across diverse domains, emphasizing scalability and user-centric automation. Smart grids extend these capabilities by integrating IoT and adaptive analytics to optimize energy efficiency and resilience against cyber-physical threats.

We categorize existing work based on the type of ML paradigm, namely supervised learning, unsupervised learning, and RL. In supervised learning, models learn from labeled data. In unsupervised learning, models learn and capture the structure of normal/regular data only~\cite{Ahmad:2017:Unsupervised:Real:Time:Anomaly:Detection:Streaming:Data, Munir:2018:DeepAnT:DL:Unsupervised:Anomaly:Detection}. Hence, in this work, methods that rely exclusively on benign data during training, also known as one-class classification, are categorized as unsupervised learning. In RL, the goal is to learn from interaction with an environment in order to secure assets. In doing so, RL agents map a state to actions via a policy function to maximize the numerical reward of the signal~\cite{Arshad:2022:Deep:RL:Anomaly:Detection:SLR}.
Comprehensive summary tables for each subsection are displayed in Tables~\ref{table: Adaptive anomaly detection in ICS applications} through~\ref{table: Adaptive anomaly detection in smart grids}.

\subsection{AAD in Industrial Control Systems (ICS)} \label{subsec:ICS}

\subsubsection{Supervised Learning} \label{subsubsec
Supervised Learning}

Nakayama et al. \cite{nakayama2019detection} proposed granger causality-based kalman filter with adaptive robust thresholding (G-KART) that learns temporal causal relationships between system components, allowing for the detection of stealthy FDIA attacks that traditional methods might miss. Authors focus on ensuring system security by analyzing interactions between components rather than relying solely on known topological relationships. Each pair of relationships is modeled using kalman filters (KF), which are updated continuously to monitor component states. The adaptive robust thresholding mechanism adjusts the anomaly detection thresholds based on the rolling median of past residuals which makes the system more resilient to noise and gradual data distribution changes. The G-KART model achieved the highest F1 score (0.141) among the tested methods. G-KART also demonstrated a significantly lower false positive rate (0.07) compared to the other models tested.

Pan et al. \cite{pan2019threshold} proposed a novel threshold-free physical layer authentication (PHY-AUC) method based on ML to enhance the security of wireless industrial CPS in mobile scenarios. Authors discuss that traditional cryptographic methods for message authentication impose high computation burdens, PHY-AUC offers an alternative with lower resource requirements. The previous PHY-AUC methods relay on fixed thresholds however, due to stability issues these iterations are ineffective in mobile environments. The authors presented a solution to this limitation by introducing a ML-based threshold-free PHY-AUC technique that authenticates as a binary classification problem. The method proposed uses channel matices as inputs for supervised ML algorithms to perform the classification of legitimate and illegitimate messages. Authors tested four ML algorithms, the best performing method was the boosted tree ensemble approach. This model contained 128 dimensional channel matrix input and reached an accuracy of 77\% in the multiple-input multiple-output (MIMO) scenario. 

Saez et al. \cite{saez2019context} proposed a context-sensitive hybrid modeling framework for cyber-physical manufacturing systems (CPMS) that aims to enhance anomaly detection and diagnosis. It combines physics-based and data-driven models, using sensor data and expert knowledge. The framework employs context-sensitive thresholds for anomaly detection and classification models for diagnosis. Applied to a CNC machine, it improved detection rates from 75\% to 94\%. The system integrates discrete and continuous states, including global operational state definitions. Physics-based models use fundamental equations, while data-driven models predict outputs with historical data. Hybrid modeling estimates variables like current and voltage, enabling accurate fault diagnosis.

Huang et al.~\cite{huang2020dynamic} proposed a dynamic game framework to model long term interactions between stealthy attackers and proactive defenders. Authors establish the foundation of the work by describing the model being comprised of two players, player 1 (user) and player 2 (defender). The user's type can be either adversarial or legitimate, which creates uncertainty for the defender with the defender's type being based on their level of sophistication (level of security awareness, detection technique, etc.). The game is structured as a multi-stage interaction, where each stage involves a sequential game with incomplete information. The defender uses an IDS to generate alerts but cannot directly identify user types. Bayesian updates refine beliefs about adversarial and legitimate users. The study derives the perfect Bayesian Nash equilibrium (PBNE) to guide strategies. Sophisticated defenders increased payoffs by 56\% and reconnaissance prevention by 41\%. The framework effectively models advanced persistent threats.

Mahdavifar et al. \cite{mahdavifar2020dennes} introduced the deep embedded neural network expert system (DeNNeS) to address the lack of explainability in DL models that are used for cyberattacks (phishing and malware). The system extracts refined rules from trained DL networks to enhance explainability and decision-making in cybersecurity applications. The framework consists of foundational elements from the matrix controlled inference engine (MACIE) algorithm (adopted from previous works), which acts as an inference justification. Authors utilized the MACIE inference justification (MIJ) algorithm and an extended version that extracts rules out of a trained multilayer deep neural network (DNN) called deep inference justification (DIJ). DeNNeS achieved a 97.2\% F1 score on a phishing dataset and 91.1\% on a malware dataset. Adaptive moment estimation (ADAM) optimized updates, enhancing classification accuracy.

Quincozes et al. \cite{quincozes2021performance} addressed the issue of cybersecurity in CPS by focusing on feature selection to improve intrusion detection across three CPS layers, perception, transmission, and application. Authors were motivated by CPS integration issues with physical components that face significant security challenges. Authors proposed a method that implements greedy randomized adaptive search procedure (GRASP) with two main phases—construction and local search. GRASP iteratively builds feature subsets and optimizes them to identify the most relevant features for intrusion detection. The GRASP method optimizes both the construction and local search phases to select relevant features across perception, transmission, and application layers. The GRASP method is tested with various classifiers (Random Forest (RF), naive Bayes (NB), J48, etc.), using datasets that are specific to CPS attack scenarios. Tested with classifiers like RF and J48, GRASP improved F1 scores by 8.29\% over traditional methods. RF achieved 99.64\% F1 in the application layer, while J48 scored 98.50\% in the perception layer. The approach enhanced CPS security across multiple layers.

Liu et al. \cite{liu2021root} introduced a novel data-driven framework for root-cause analysis of anomalies in complex CPS using a spatiotemporal graphical modeling approach based on symbolic dynamics. The framework focuses on discovering and representing causal interactions among subsystems. The authors formulated the root-cause analysis problem as a minimization problem using an inference-based metric and proposed two approximate solutions, sequential state switching (S3) and  artificial anomaly association (A3). The effectiveness of these methods was validated using synthetic data and the Tennessee eastman process (TEP) dataset. The S3 method analyzed patterns of subsystem interactions sequentially to identify potential root causes and the A3 method treats root-cause analysis as a classification problem using a neural network (NN) model. Using synthetic and real datasets, the results showed that S3 and A3 could effectively identify faulty nodes, with S3 having a 100\% recall, precision, and F1 score, while A3 achieved a recall of 96.7\%, precision of 90.6\%, and an F1 score of 93.6\%.

Althobaiti et al.~\cite{althobaiti2021intelligent} introduced a novel cognitive computing-based IDS for industrial CPS that leverages AI to address security issues. The approach uses a binary bacterial foraging optimization (BBFO) for feature selection and a gated recurrent unit (GRU) for classification. BBFO is inspired by the foraging behavior of bacteria, where bacteria move towards optimal solutions in the search space. This approach aims to reduce the dimensionality of the dataset, improving computational efficiency and model accuracy.  Nesterov-accelerated adaptive moment estimation (NADAM) is employed to optimize the hyperparameters of the GRU. Applied to industrial datasets, the system achieved 98.45\% detection accuracy. The method combines high precision with reduced computational costs.

Vavra et al.~\cite{vavra2021adaptive}, presented a comprehensive approach to anomaly detection in ICS using ML algorithms. Recognizing the critical role of ICS in modern society and the increasing risks posed by cyberattacks, the authors develop an AAD system designed to address key challenges, including the detection of unknown attacks, scalability, adaptability, high false alarm rates, and computational complexity. The system integrates artificial neural networks (ANN), long short-term memory networks (LSTM), isolation forest (IF), and one-class support vector machines (OCSVM), each adapted for semi-supervised learning. This approach enables the system to identify anomalies by learning from normal operational data and flagging deviations as potential cyberattacks. The use of semi-supervised learning is particularly important for detecting unknown attacks that may not be present in the training data. The paper also emphasizes the importance of data preprocessing and feature selection, noting that issues like missing values, feature scaling, and high dimensionality of data can significantly impact the performance of ML models. Techniques such as principal component analysis (PCA) are employed to reduce the dimensionality of the dataset while retaining the most critical information, thereby enhancing the system's efficiency and scalability. The paper concludes that the proposed system effectively balances detection accuracy with computational efficiency, particularly when the IF algorithm is optimized using genetic algorithms (GA). The system balances detection accuracy and computational demands, achieving strong performance metrics.

Alohali et al. \cite{alohali2022artificial} proposed an AI-enabled multimodal fusion-based intrusion detection system (AIMMF-IDS) that employs an improved fish swarm optimization (IFSO-FS) technique for feature selection by using the Levy Flight concept to enhance searching capability and avoid local optima problems. The system integrates recurrent neural network (RNN), bi-directional LSTM (Bi-LSTM), and DBN models for multimodal fusion. Applied to NSL-KDD and CICIDS datasets, it achieved precision, recall, and F1 scores above 95\%. The model addresses intrusion detection challenges with robust performance.

Ibor et al. \cite{ibor2022novel} presented a novel hybrid approach to predict cyberattacks on CPS communication networks. The proposed method leverages a bio-inspired hyperparameter search technique to improve the structure of a deep neural network (DNN). The GA operates by generating an initial population of neural network (NN) structures (chromosomes), evaluating their performance using a fitness function, and then applying selection, recombination, and mutation to generate improved structures over multiple generations. Tested on CICIDS2017 and UNSW-NB15 datasets, the method achieved 99.81\% and 80.10\% accuracy, respectively. The approach optimizes neural network architecture for better attack prediction with minimal false positives.

Liu et al. \cite{liu2022intrusion} investigated enhancing cybersecurity for maritime transportation systems (MTS) using IoT technology. Authors proposed a federated learning-based IDS called ``FedBatch" that employs a hybrid convolutional neural network (CNN) multilayer perceptron (CNN-MLP) model. The model addresses the issue of maintaining data privacy and handling the straggler problem (delays in data/model updates due to unstable communication in maritime settings). It adapts to unstable communication environments using dynamic aggregation methods. Tested on independent and identically distributed (IID) and non-IID datasets, FedBatch achieved 88.1\% accuracy, outperforming traditional federated learning methods.

Kure et al.~\cite{kure2022asset} presented a unified approach to cybersecurity risk management (CSRM) for CPS by integrating fuzzy set theory and ML classifiers. The focus is on predicting risk types, assessing asset criticality, and evaluating the effectiveness of existing controls. Fuzzy set theory is applied to assess the criticality of CPS assets, considering security factors such as confidentiality, integrity, availability, accountability, and conformance. Multiple ML classifiers, including k-nearest neighbors (k-NN), decision tree (DT), RF, and NB, are used to predict different risk types like DoS, cyber espionage, and crimeware. DT achieved 93\% accuracy in predicting risk types. Fuzzy logic improves asset assessment, supporting proactive risk management.

Shi et al. \cite{shi2023lstm} proposed a ML-driven online side-channel monitoring approach that utilizes an LSTM autoencoder (AE) for detecting unintended alterations in the additive manufacturing (AM) process. Authors were motivated by the growing vulnerabilities in AM due to potential cyber-physical attacks. Traditional monitoring methods fail to detect internal alterations in AM parts, which can severely compromise the functionality and mechanical properties of the product. Both supervised and unsupervised monitoring schemes were implemented, with experimental validation conducted on a fused filament fabrication (FFF) platform equipped with accelerometers. The AE, consisting of an encoder and a decoder, where the encoder reduces high-dimensional input data into a compact latent representation using an encoding function implemented with LSTM layers, which effectively capture temporal dependencies in sequential sensor data. The decoder reconstructs the original input from the latent representation using a decoding function. The reconstruction error quantifies the difference between the original input and the reconstructed data, serving as a measure of how well the model represents the input. Two cases are considered for this investigation, Case 1 involves inserting a void into the design geometry at the STL stage, while Case 2 alters layer thickness at the slicing stage. In supervised monitoring, the proposed method achieved high F1 scores of 0.95 in Case 1 and 0.96 in Case 2, outperforming traditional methods. In unsupervised monitoring, the LSTM-AE with OCSVM exponentially weighted moving average (OCSVM-EWMA) demonstrated a low false alarm rate (0.09 in Case 1 and 0.03 in Case 2) and fast attack response times (4.6 and 7.4 samples, respectively).

Intriago et al. \cite{intriago2023real} introduced a Hoeffding adaptive tree (HAT) classifier combined with an instance selection algorithm to detect cyber and non-cyber contingencies in real-time for cyber-physical power systems (CPPS). Authors are motivated based on challenges presented in continuous monitoring of CPPS through wide-area monitoring, protection, and control (WAMPAC) systems that handle high-velocity and unbounded data streams from devices like phasor measurement units (PMUs). Authors addressed this issue by creating a streaming learning classification system that adapts to evolving data streams. They proposed a three-stage instance selection process involving reordering, resetting outdated data, and dynamic window size selection. The classifier outperformed existing methods in six case studies, achieving over 99\% accuracy on multiclass datasets. The HAT stream instance selection (HAT+SIS) classifier achieved over 99\% accuracy in the multiclass dataset, maintained this accuracy even as the data evolved, and  outperformed HAT drift detection method (HAT+DDM) and Hoeffding tree (HT+DDM). These methods experienced significant accuracy drops around 4,000 instances. HAT+SIS also processed approximately 5,000 instances in 43 seconds and maintained a stable model size of around 196 KB, demonstrating both efficiency and scalability for real-time applications.

Alshammari~\cite{alshammari2023design} proposed a Bayesian neural network (BNN) architecture for evaluating the statistical features of DoS attacks. In the investigation, the NSL-KDD dataset is used which is a comprehensive network security dataset. The paper emphasizes selecting features with the highest predictive power while minimizing redundancy which is established with a heatmap of feature mean values. The BNN applies the local markov property to forecast event probabilities and represent probabilistic relationships between different variables using a directed acyclic graph. The BNN is constructed with multiple nodes, each of the nodes represents a random variable with the edges showing conditional dependencies between variables. The preprocessed data is trained using the BNN to learn patterns and relationships in the network traffic. The author states that the BNN demonstrated a testing accuracy of 97.5 \% with other metrics highlighting that the BNN was capable of generalizing well. When compared to accuracies from DT (78\%-89\%) and ANN (87\%-89\%), the BNN outperforms these traditional networks in terms of identifying anomalies.

Wang et al. \cite{wang2024real} introduced a novel integrated ML and DL approach for real-time attack detection and identification. The authors present a two-stage solution involving a OCSVM for detecting whether a CPS is under attack, followed by a pairwise self-supervised long short-term memory (PSLSTM) model for identifying the attack type. The method is designed to detect known attack types and discover unknown new attacks. The two-stage learning approach begins with the OCSVM model that is trained exclusively on data representing the normal state of the CPS. The OCSVM model functions by defining a hyperplane that separates normal data points from potential outliers. The second stage begins once the attack is detected. PSLSTM is used to identify the specific attack type and consists of several LSTM networks that effectively convert a multi-class problem into a binary classification task. The number of pairwise models to be trained is calculated from a given \(\mathbf{K}\) known attack types and \(\frac{\mathbf{K}(\mathbf{K}-1)}{2}\) LSTM models. The OCSVM-PSLSTM method demonstrated high-quality performance with real-time attack detection and identification by achieving an average accuracy of 99.7\% for identifying known attack types and 97.2\% with discovering unknown attacks.

\subsubsection{Unsupervised Learning} \label{subsubsec:ICS Unsupervised Learning}
Mitchell et al. \cite{mitchell2013survivability} developed a mathematical model to assess the survivability of mobile CPS (MCPS) using dynamic voting-based intrusion detection. They employed a stochastic Petri net (SPN) model to analyze trade-offs between energy conservation and intrusion tolerance. The system dynamically adapts to changing system states and environmental conditions. As the proportion of compromised nodes or energy levels change, the model adjusts the intrusion detection interval  \(T_{IDS}\) and the number of detectors in real time to maintain optimal system performance. The dynamic voting-based technique continuously updates its parameters without requiring pre-labeled data, allowing the system to respond to attacks and energy constraints as they occur. The results demonstrate that there is an optimal \(T_{IDS}\) that maximizes the mean time to failure (MTTF) of the MCPS by balancing energy consumption and intrusion tolerance. Authors found that an optimal value of \(T_{IDS}\)=160s with 5 detectors provided the highest MTTF for their reference system. The model's predictions were validated through simulations, showing a close match with theoretical results, with only a 4.60\% to 7.64\% mean percentage error.

Meria et al. \cite{meira2020performance} examined six unsupervised algorithms; AE, one-class nearest neighbor, IF, one-class $\mathcal{K}$-means, one-class scaled convex hull, and OCSVM, on two public datasets (NSL-KDD and ISCX) for anomaly detection in cybersecurity. Data was preprocessed by normalization using $\mathcal{Z}$-score and Min Max. The one-class scaled convex hull achieved the highest AUC on the NSL-KDD dataset with an AUC value of 85.30\%. The one-class nearest neighbor performed best on the ISCX dataset with an AUC of 95.20\%. Overall, the one-class nearest neighbor, one-class scaled convex hull, and OCSVM demonstrated the best performance across both datasets.

Xi et al.~\cite{xi2023adaptive} proposed the adaptive-correlation-aware unsupervised DL (ACUDL) which addresses the challenge of detecting anomalies in high-dimensional, noisy, and unlabeled data. The core innovation of ACUDL is its use of a dynamic graph structure to represent and update the implicit correlations among data points, which is critical for accurately capturing the underlying relationships in CPS data. ACUDL begins by constructing an initial directed graph using the KNN algorithm, where each node in the graph represents a data point, and edges represent the correlations between these points. This graph is then dynamically updated during training to reflect changes in the data, using an adaptive mechanism that adjusts the graph structure based on the training loss. The model also incorporates a dual-AE (D-AE) framework that separately encodes the original non-correlation features, the correlation features extracted from the graph, and a decoder. These features are then fused and passed through a gaussian mixture model (GMM) to estimate the anomaly energy, which is used to detect anomalies.  Through extensive experiments on various CPS datasets, including scenarios like smart healthcare systems (SHS) and intelligent cruise control systems (ICCS), the authors demonstrate that ACUDL significantly outperforms existing DL-based anomaly detection methods. Experiments on CPS datasets demonstrated its superiority, achieving AUC scores of 73.3\% and 87.5\%, with strong F1 scores and precision. ACUDL effectively handles high-dimensional, noisy data.

Cai et al. proposed an adaptive distributed denial-of-service (DDoS) mitigation scheme for software-defined networking (SDN)~\cite{cai2023adam}. Authors proposed the adaptive DDoS attack mitigation (ADAM) scheme that combines information entropy and unsupervised anomaly detection methods to detect both known and unknown DDoS attacks in software-defined CPS (SD-CPS). Motivated by the increasing vulnerability of CPS to DDoS attacks due to insecure or outdated components, the authors highlight the limitations of traditional defense mechanisms that rely on static thresholds or predefined attack signatures. ADAM operates through three stages: nominal, detection, and mitigation. In the nominal stage, a ``nominal profile'' of normal network traffic is created by sampling traffic and calculating entropy vectors for features such as IP addresses and port numbers. ADAM achieves a high accuracy of 99.13\% on average in mitigating various DDoS attacks, with a significantly reduced false-positive rate compared to existing methods. This suggests that ADAM is a scalable and effective solution for SD-CPS environments.

Table \ref{table: Adaptive anomaly detection in ICS applications} summarizes key attributes of the papers reviewed in Section \ref{subsec:ICS}.

\onecolumn
\scriptsize
\begin{longtable}{|p{2cm}|p{2cm}|p{1.5cm}|p{1.5cm}|p{1.5cm}|p{3cm}|p{3cm}|}
\caption{Summary of AAD methods in ICS applications.}
\label{table: Adaptive anomaly detection in ICS applications} \\
\hline
\textbf{Reference} & \textbf{Model} & \textbf{Algorithm} & \textbf{Dataset} & \textbf{Attack} & \textbf{Strengths} & \textbf{Weaknesses} \\
\hline
\endfirsthead
\multicolumn{7}{c}{\textit{Table \thetable{} -- continued from previous page}} \\
\hline
\textbf{Reference} & \textbf{Model} & \textbf{Algorithm} & \textbf{Dataset} & \textbf{Attack} & \textbf{Strengths} & \textbf{Weaknesses} \\
\hline
\endhead
\hline \multicolumn{7}{r}{\textit{Continued on next page}} \\
\endfoot
\hline
\endlastfoot
Mitchell et al.~\cite{mitchell2013survivability} 2013 & Unsupervised & SPN & Simulation of MCPS & Node capture and bad data injection & Comprehensive mathematical model to balance energy consumption and intrusion detection & Not demonstrated that model generalizes well on real-world data \\
Nakayama et al.~\cite{nakayama2019detection} 2019 & Supervised & G-KART & IEEE 33-bus power & FDI attacks & Adaptive robust thresholding mechanism & Tested in simulated environment \\ 
Pan et al.~\cite{pan2019threshold} 2019 & Supervised & DT, SVM, KNN, and boosted trees & Real-world, large-scale channel measurement campaign from 4 real industrial sites & Spoofing attacks & System can adapt to dynamic environments and significantly improve authentication accuracy & Offline learning \\ 
Saez et al.~\cite{saez2019context} 2019 & Supervised & Context-sensitive adaption, SVM & Sensor data, CNC machine & Anomalies only & Adaptive system & Not tested on attacks \\ 
Huang et al.~\cite{huang2020dynamic} 2020 & Supervised & PBNE, dynamic game & TEP, Time-ordered simulation & APT, user-based, finite options & Dynamic game framework that offers proactive and adaptive approaches to enhance security & Framework's complexity might limit its scalability to larger systems \\ 
Meria et al.~\cite{meira2020performance} 2020& Unsupervised & AE, one-class nearest neighbor, one-class K-Means, IF, one-class scaled convex Hhull, and OCSVM & NSL-KDD and ISCX & DoS, R2L, U2R, probe, brute force & Shows unsupervised methods can effectively detect unknown attacks & Offline learning \\ 
Mahdavifar et al.~\cite{mahdavifar2020dennes} 2020 & Supervised & DENNES, MACIE & UCI phishing websites and Android malware & Website phishing and Android malware & Efficiently explains the causes of cyber threats and outperforms other explainable algorithms (RF) & Lacks performance of basic DNN \\ 
Quincozes et al.~\cite{quincozes2021performance} 2021 & Supervised & GRASP & SWaT-CPS, NSL-KDD, WSN-DS,CICIDS2017 & CPS injection, flooding, grayhole, blackhole, TDMA, ARES DDoS, LOIC DDoS, PortScan, R2L, U2R & Model generalizes well across datasets & Relies on offline feature selection \\
Liu et al.~\cite{liu2021root} 2021 &  Un-supervised (S3) and supervised (A3) & S3, A3 & TEP & faults within CPS & Effective at root-cause analysis for both pattern-based and node-based anomalies & S3 method may be computationally intensive \\ 
Althobaiti et al.~\cite{althobaiti2021intelligent} 2021 & Supervised & BBFO, GRU & NSL-KDD 2015 and CICIDS 2017 & General anomalies (related works does mention replay attacks) & BBFO improved system efficiency, GRU’s  makes it suitable for range of CPS & Requires hyperparameter optimization \\ 
Vavra et al.~\cite{vavra2021adaptive} 2021 & Unsupervised & LSTM, IF, and OCSVM & Secure Water Treatment (SWaT), ICS network communication & General cyber-attack, anomaly detection & System effectively handles high-dimensional data using PCA & System struggles with interpreting the specific nature of the detected anomalies \\ 
Alohali et al.~\cite{alohali2022artificial} 2022 & Supervised & IFSO that combines RNN, Bi-LSTM, and DBN & NSL-KDD 2015 and CICIDS 2017 & General intrusion detection & Feature selection efficiently reduces data dimensionality & Offline learning \\ 
Ibor et al.~\cite{ibor2022novel} 2022 & Supervised & Bio-inspired deep feedforward, modified genetic search & CICIDS2017 and UNSW-NB15 & Brute Force, Heartbleed, Botnet, DoS, DDoS, Web Attack, Infiltration, Analysis, Backdoor, Exploits, Fuzzers, Generic, Reconnaissance, Shellcode, Worms & Bio-inspired hyperparameter search allows dynamic optimization & Offline model \\ 
Liu et al.~\cite{liu2022intrusion} 2022 & Supervised & FedBatch, CNN-MLP & NSL-KDD & 40 attack types divided into five categories: Normal, DoS, Probe, U2R and R2L & Resilience in handling non-IID data & Assumes a uniform distribution of stragglers (might be dynamic in real environments) \\ 
Kure et al.~\cite{kure2022asset} 2022 & Supervised & DT, NB, RF, KNN & VCDB & Crimeware (R1), cyber espionage (R2), denial of service (R3), everything else (R4), lost and stolen assets (R5), miscellaneous errors (R6), payment card skimmers (R7), point of sale (R8), privilege misuse (R9), and web applications (R10) & Integration of fuzzy set theory and machine learning for proactive risk prediction is innovative. & Study does not focus on streaming data \\ 
Shi et al.~\cite{shi2023lstm} 2023 & Supervised (classification) and unsupervised (novelty detection) & LSTM-AE & Accelerometer data from FFF platform & Void in the STL design, alteration in slicing stage & Effectively captures temporal dependencies & Test parts used in the experiment have simple designs \\ 
Intriago et al.~\cite{intriago2023real} 2023 & Supervised & HAT with a novel instance selection algorithm & multiclass industrial control system cyber-attack dataset & Line maintenance, short-circuit faults, remote tripping command injection attacks, relay setting change attacks, and data injection attacks & Adapts effectively to evolving data streams & Need for tuning hyperparameters and the potential alteration of the temporal distribution due to instance reordering \\ 
Alshammari ~\cite{alshammari2023design} 2023 & Supervised & BNN & NSL-KDD (network security dataset) & DoS & Higher performance compared to traditional ML methods & Performance depends on feature selection \\ 
Xi, et al.~\cite{xi2023adaptive} 2023 & Unsupervised & ACUDL, dynamic graph update, D-AE & ECG5000, Arrhythmia, Satellite, and CIFAR-10 & Anomaly detection, general cyber attack & Systems uses a D-autoencoder that uses correlated and non-correlated features  & Might not be scalable \\ 
Cai et al.~\cite{cai2023adam} 2023 & Unsupervised & ADAM , KNN  & MAWI Working Group Traffic & Volumetric DDoS attacks & High accuracy, low FPR, designed to be scalable & Vulnerability of SDN, only focused on volumetric attacks \\ 
Wang et al.~\cite{wang2024real} 2024 & Unsupervised (OCSVM) and supervised (PSLSTM) & OCSVM, PSLSTM & Lab experiment, 12 raspberry PIs & Cut, virus, trojan, scan, intrusion, and heat & Detects both known and unknown attacks & Requires substantial computational resources \\ 
\end{longtable}
\twocolumn
\normalsize

\subsection{AAD in Vehicle Networks} \label{subsec:Vehicle}


\subsubsection{Supervised Learning} \label{subsubsec:vehicle Supervised Learning}

van Wyk et al.~\cite{Van:2019:Real:Time:Anomaly:Detection:Automated:Vehicles} developed an anomaly detection approach that combines CNNs and a Kalman filter (KF) with a $\chi^2$ detector to identify anomalous behavior in connected and automated vehicles (CAVs). They use a sliding window approach to focus the analysis on the latest observations. Here, the input to the CNN module is a stream of images from a continuous feed of raw sensor data during a CAV trip. They trained a separate CNN model per sensor using labeled images. They used the ``OR'' logical operation of the outcomes in each of the sensors to determine if anomalous readings are detected across all sensors. To improve further on detection, the output of the CNN feeds an adaptive KF with a $\chi^2$ detector for further examination for anomaly detection. KF has prediction and update phases. In the prediction phase, the KF produces estimates of the the time series including their uncertainties. Once the predictions of the next samples are calculated, estimates are updated using a weighted average with greater weights to estimate with greater certainty. KF works recursively. They tested their approach using data from the Safety Pilot Model Deployment (SPMD) program~\cite{Bezzina:2014:Safety:Pilot:Model:Deployment} that demonstrate CAVs in action. In particular, they focused on analyzing time series from three sensors: in-vehicle speed, GPS speed, and in-vehicle acceleration for a vehicle with a trip length of 2,980 seconds. As the original data contains no anomalies, authors inject synthetic anomalies, including instant (simulated as a random Gaussian variable), constant (a temporarily constant observation that is different from the normal), gradual drift (by linearly increasing a set of values to the base value of the sensors), and bias (a temporarily constant offset from the baseline sensor readings). 

Feng et al.~\cite{Feng:2020:Efficient:Drone:Hijacking:XGBoost} proposed an efficient drone hijacking detection method that consumes inertial measurement unit (IMU) (i.e., gyroscope and accelerometer) and GPS (i.e., longitude and latitude) data. The proposed method used the eXtreme Gradient Boosting (XGBoost) algorithm to mine the relationship between IMU and GPS data using real-time data samples to decide if the drone has been hijacked or not. The model is first trained offline where parameters are optimized using a GA. In the deployment state, the same training parameters are used to update the model onboard. Experiments are conducted on a real quadcopter with an oﬀ-the-shelf multi-core embedded board and an autopilot sensor board. Prediction correctness in each sample time was reported as $96.3 \%$ and $100\%$ in hijack and normal scenarios. As the proposed model is deployed in online fashion it can achieve $100\%$ detection correctness just after 1 second after the hijack starts. 

Alsulami et al.~\cite{Alsulami:2023:Security:Autonomous:Vehicle:Transfer:Learning} proposed an intelligent intrusion detection systems (IIDS) for autonomous vehicle-cyber physical systems (AV-CPS) that focuses on transfer learning. Specifically, the proposed method used eight pre-trained CNNs, including InceptionV3, ResNet-50, ShuffleNet, MobileNetV2, GoogLeNet, ResNet-18, SqueezeNet, and AlexNet. By leveraging pre-trained models to enhance the performance of IIDS, authors target the detection of anomalous communications in the controller area network (CAN) bus having an effect on the connected physical components of AVs. Authors' simulation setup include a self-driven car system consisting of a lead and an ego vehicle (self-driving car). In an ideal situation, the ego vehicle maintains its distance from the lead vehicle using the adaptive cruise control system. Authors simulate a CAN communication network environment using Simulink and produced a dataset consisting of (1) position of the ego vehicle, (2) velocity of the ego vehicle, (3) position of the lead vehicle, and (4) velocity of the lead vehicle. Note that for the data be processed by the CNNs, time series representations of captured signals are tranformed into a 2-dimentional representation (i.e., images). They found that GoogLeNet performed best achieving $99.47\%$ on F1 score metric.

\subsubsection{Reinforcement Learning} \label{subsubsec:vehicle reinforcement Learning}


Mowlka et al.~\cite{Mowla:2020:AFRL:Adaptive:Federated:RL:Jamming} proposed an adaptive federated reinforcement learning-based jamming attack defense strategy to protect flying ad-hoc networks (FANETs), i.e., a decentralized communication network unmanned aerial vehicles (UAVs). They focused on a model-free Q-learning mechanism with adaptive exploration-exploitation epsilon greedy policy, directed by an on-device federated jamming detection mechanism. Q-learning learns a policy maximizing total rewards based on  trial-and-error. That is, a UAV client receives a negative reward if it is moved closer to the jammer location. Here, an epsilon-greedy policy is used to balance the outcome of exploration and exploitation opportunities. They showed that the proposed adaptive federated RL-based approach performed better spatial retreat defense strategies.

Table~\ref{table: Adaptive anomaly detection in vehicular applications} summarizes key attributes of the papers reviewed in Section~\ref{subsec:Vehicle}.

\begin{table*}[htp!]
\centering
\caption{Summary of AAD methods in vehicular applications.}
\label{table: Adaptive anomaly detection in vehicular applications}
\small 
\begin{adjustbox}{max width=\textwidth}
\begin{tabular}{|p{2.5cm}|p{1.5cm}|p{1.5cm}|p{2cm}|p{2.5cm}|p{3.5cm}|p{3.5cm}|}
\hline
{Reference} & {Model} & {Algorithm} & {Dataset} & {Attack} & {Strengths} & {Weaknesses} \\
\hline
van Wyk et al.~\cite{Van:2019:Real:Time:Anomaly:Detection:Automated:Vehicles} 2019 &  Supervised & CNN, KF & Public real CAV data & Instant, constant, gradual, bias drifts  & CNN and KF approach combines strengths & Tests of simulated attacks  \\

Feng et al.~\cite{Feng:2020:Efficient:Drone:Hijacking:XGBoost} 2020 & Supervised & XGBoost & Real prototype quadrotor drone & GPS spoofing attacks  & Model further trained onboard & Limited number training scenarios  \\

Mowla et al.~\cite{Mowla:2020:AFRL:Adaptive:Federated:RL:Jamming} 2020 & RL & Q-learning & Simulated FANET topology & Constant, random, and reactive jamming & Reduce the number of route jammer location hop counts  & Small-scale simulation  \\

Alsulami et al.~\cite{Alsulami:2023:Security:Autonomous:Vehicle:Transfer:Learning} 2023 & Supervised & CNN  & AV simulation & False data injection  & Transfer learning allows to use pre-trained models  & Limited to false data injection attacks  \\
\hline
\end{tabular}
\end{adjustbox}
\end{table*}


\subsection{AAD in Power Grids} \label{subsec:Power Grid}


\subsubsection{Supervised Learning} \label{subsubsec:power grid Supervised Learning}

Cui et al.~\cite{Cui:2021:Source:Authentication:Distributed:Synchrophasors:Microgrids} proposed a hybrid approach combining self-adaptive mathematical morphology (SAMM) and time frequency (TF) techniques to authenticate source information on distribution synchrophasors (DS) within microgrids at near-range locations. Their proposed method is intended to deter ``source ID mix'' data spoofing attacks on DS as they threat to the security of the power grid. This attack can manipulate source information of DS without changing the measurement values. Manipulating the source information on DS has an effect on critical synchrophasor-based control and applications including wide-area damping and control and disturbance localization. SAMM allows for adaptive regulation of synchrophasors variations which represent local environmental characteristics. In addition, TF mapping is used to extract frequency-related features from the regulated synchrophasors variation. They used a RF classifier to correlate the extracted signatures with the source information based on the TF analysis. Their proposed approach consists of three steps including (1) a high-pass filter to extract frequency variation on the original DS data; (2) an integrated SAMM-TF used to extract useful features from the synchrophasors variation; and (3) the integration with a random forest classifier for source integration. The SAMM approach adaptively preserves significant peaks in frequency so that distinctive signatures can be extracted for source authentication. They validated their results using distribution synchrophasors from multiple small geographical scales validating the proposed methodology.

Khan et al.~\cite{Khan:2021:Privacy:Preserving:IDS:Power:Networks} proposed a privacy-conserving based intrusion detection (PC-IDS) for contemporary smart power systems (SPNs) using a hybrid ML approach. The proposed framework consist of two main components: data pre-processing and intrusion detection. Data pre-processing entails attribute/feature mapping, reduction, and normalization. This allows to process diverse types of attributes such as numerical and categorical features. The intrusion detection module comprises stages of training and detection. Specifically, it uses a PNN and consumes several types of regular and malicious patterns to enhance classification performance. Particle swarm optimization (PSO) is used to select the hyperparameters of the PNN model. The performance of the proposed framework is evaluated in two commonly available datasets: power system~\cite{Morris:2013:ICS:Attack:Datasets} and UNSW-NB15~\cite{Moustafa:2015:UNSW:Dataset} datasets. Their experimental evaluation shows the effectiveness of the proposed framework to protect data from SPNs and determine anomalous behavior in terms of traditional evaluation metrics. 

Jiao et al.~\cite{Jiao:2022:Cyberattack:Resilient:Forecasting:Adaptive:Robust:Regression} proposed a cyberattack-resilient load forecasting approach based on robust regression, i.e., adaptive least trimmed squares (ALTS)~\cite{Bacher:2016:Adaptive:Robust:Regression:ALTS}. Authors assume that adversaries alter load entries in the training data so that the estimated regression coefficients become inaccurate resulting in forecasts that may lead to poor decision making. They used two different attack types: random and ramping attacks. In random attacks, a randomly selected proportion of the training data is scaled by a random factor following, for example, a normal distribution. In ramping attacks, many single attack intervals are injected parametrized by starting attack point and length. They injected these attacks to alter the GEFCom2012 dataset~\cite{Hong:2014:Global:Enery:Forecasting:Dataset}. Their robust approach focused on estimating robust estimators for the coefficients in the regression models based on M-estimation~\cite{Huber:1992:Robust:Estimation:Location:Parameter}---a generalization of the maximum likelihood estimation. The M-estimator is obtained through the iterative re-weighted least squares (IRLS) algorithm. The core of their adaptive algorithm is based on least trimmed squares (LTS) which is a robust alternative to ordinary LS when dealing with linear regression problems. LTS minimizes the sum of square residuals over a subset of the whole datapoints by excluding a proportion of $p$ data points whose residuals are the largest in magnitude. Thus $p$ is a tuning parameter that leads to the ALTS method. This provides robustness to potential outliers. They compute goodness of fit using mean absolute percentage error. Five methods including, LS, M-Huber, M-bisquare, L$_{1}$, and ALTS ere used to fit the model and perform forecasting in the validation dataset. A comparison analysis using the GEFCom2012 dataset suggests that the ALTS method is robust against to random and ramping attacks both when the proportion of attack data is high and the robustness does not decrease as attack data proportion increases. 

Ding et al.~\cite{Ding:2022:Data:Driven:Situational:Awarness:Power:systems} proposed a data-driven security situational awareness framework to secure power systems. Their proposed framework focused on an adaptive honeypot architecture for capturing system logs and network traffic in distributed fashion. Specifically, they deployed 10 honeypots around the world by simulating industrial control devices in power systems to lure attackers. Their deployed honeypot architecture includes multiple industrial protocols including Modbus, Siemens S7-Comm, Guardian$\_$ast, Kamstrup$\_$382, Bacnet, Http and Ipmi. The honeypot architecture was deployed using the Alibaba cloud environment. Security incidents are detected by modeling the captured attack traffic from honeypots and the security scanning system. In doing so, they constructed a security situation graph based on traffic logs by modeling IP instances as nodes in a graph. They annotate the nodes in the graph using word2vec~\cite{Mikolov:2013:Efficient:Estimation:Word:Representations:Word2Vec} by processing attributed information of IPs. Their detection algorithms leverage a graph convolutional network (GCN) to detect malicious IPs. Under this approach, the GCN learns the features of malicious IPs in a supervised fashion. They performed experiments to evaluate performance of TCP SYN probe in their deployed scanning system. They tested two attack configurations: single port scanning and multi-port scanning. 

Table~\ref{table: Adaptive anomaly detection in power grids} summarizes key attributes of the papers reviewed in Section~\ref{subsec:Power Grid}.
\begin{table*}[htp!]
\centering
\caption{Summary of AAD methods in power grid applications.}
\label{table: Adaptive anomaly detection in power grids}
\begin{adjustbox}{max width=\textwidth}
\begin{tabular}{|p{2.5cm}|p{1.5cm}|p{1.5cm}|p{2cm}|p{2.5cm}|p{3.5cm}|p{3.5cm}|}
\hline
{Reference} & {Model} & {Algorithm} & {Dataset} & {Attack} & {Strengths} & {Weaknesses} \\
\hline
Cui et al.~\cite{Cui:2021:Source:Authentication:Distributed:Synchrophasors:Microgrids} 2021 & Supervised & SAMM and TF & Photovoltaics nodes & FDIA & Adaptation to synchrophasors variations & Lack of continual learning \\

Khan et al.~\cite{Khan:2021:Privacy:Preserving:IDS:Power:Networks} 2021 & Supervised & PNN & Phasor data concentrator and network data & FDIA & PNN train faster than MLP & Lack of continual learning \\

Jiao et al.~\cite{Jiao:2022:Cyberattack:Resilient:Forecasting:Adaptive:Robust:Regression} 2022 & Supervised & ALTS & Global energy forecasting competition & FDIA & Robust against random and ramping attacks & Only focus on linear regression  \\

Ding et al.~\cite{Ding:2022:Data:Driven:Situational:Awarness:Power:systems} 2022 & Supervised & GCN & TCP SYN probes & Information leakage, security vulnerability, DoS attack, APT attack, SQL injection, Malware infection, remote attack & Graph modeling that fuses inside and outside cyber threat incidents & Lack of continual learning \\
\hline
\end{tabular}
\end{adjustbox}
\end{table*}

\subsection{AAD in the Internet of Things (IoT)} \label{subsec:IoT}


\subsubsection{Supervised Learning} \label{subsubsec:IoT Supervised Learning}

Li et al.~\cite{Li:2019:System:Statistics:Learning:IoT} introduced a statistical learning based anomaly detection technique that monitors the operation of IoT devices to detect possible cyberattacks and malicious activities. In particular, they used time series derived from system statistics including CPU usage, memory consumption, and network throughput, among others, to model  normal behavior. To do so, they simulated a real IoT system operation made of 12 beagle bone black (BBB) connected to a router in a LAN topology. In particular, they ran a program that samples random data with a fixed interval and processes a variety of signal processing, storage, compression, and transmitting operations. They injected simulated cyberattacks including unauthorized access, port scan, virus, and flood. They trained LR, NN, and recurrent neural network (RNN) classifiers to predict normal system behaviors. These models were trained using different window sizes showing that there is a threshold for which the MAE starts getting diminishing returns. Their results showed that the NN perform the best at the expense of higher computational complexity. The thresholding for deciding if a particular data sample is an anomaly or not is performed through local outlier factor (LOF), cumulative statistics thresholding (CUSUM), and adaptive online thresholding (AOT). They showed that the method using AOT perform better than LOF and CUSUM based on F1 score evaluation metric. 

Gopalakrishnan et al.~\cite{Gopalakrishnan:2020:DL:Data:Offloading:Mobile:Edge:Computing:System} introduced a deep learning based traffic prediction framework with offloading mechanism and cyberatack detection (DLTPDO-CD). Their proposed approach is composed of three major processes involving traffic prediction, data offloading, and attack detection. In doing so, it includes first a BiLSTM based traffic prediction process to enable proficient data offloading. Considering a double LSTM is meant to enhance learning long-term dependencies as it enhances the accuracy of the detection. Second, an adaptive sampling cross entropy technique (ASCE) is incorporated to maximize network throughput by deciding offloading users from the network. Lastly, for the detection of cyberattacks in mobile edge computing, they used a DBN optimized using a barnacles mating optimizer (BMO). Their approach was tested on simulated data showing better performance over compared methods under different dimensions.

Bibi et al.~\cite{Bibi:2022:Deep:AI:Cyber:Threat:Analysis:IIoT} proposed an efficient and self-learning autonomous multivector threat intelligence and detection mechanism to proactively defend IIoT networks. Their approach used a convolutional LSTM2D (Cu-ConvLSTM2D mechanism) being highly scalable with self-optimized capabilities to detect diverse and dynamic variant of IIoT threats. Cu-ConvLSTM2D is a recurrent layer similar to LSTM, but the internal matrix multiplication is exchanged with convolution operations. They evaluated their proposed framework on a Kitsune surveillance network intrusion dataset~\cite{Mirsky:2018:Kitsune} comprising 21 million instances of varying attack patterns and prevalent threat vectors. The proposed technique outperforms current contemporary DL-driven architectures and existing benchmarks.

Yazdinejad et al.~\cite{Yazdinejad:2023:Secure:Fuzzy:Blockchain:Framework} proposed a novel design and implementation of a secure and intelligent fuzzy blockchain framework. Their proposed framework focus on three layers: IoT, blockchain, and intelligent fuzzy layers. The IoT layer hanldes smart devices that communicate with each other in the blockchain environment. The blockchain layer handles IoT device management ensuring safe channels for the transmission of data and transactions between IoT devices. In their intelligent fuzzy layer, they used a threat detection in the blockchain layer. Their threat detection module is based on a fuzzy DL model that uses a fully connected network based on fuzzy neurons to output the aggregation of classification results of several classifiers. In conjuction, an adaptive neuro-fuzzy inference system (ANFIS) model is used to design an optimal fuzzy system for threat detection in IoT networks. The ANFIS model estimates input membership functions and output modified membership functions. Finally, a fuzzy control system module leverages previous inputs in both blockchain and IoT layers to feed a fuzzy control system module to arrive at complex decision making. Their proposed framework was tested for threat detection in the Ethereum blockchain~\cite{Jung:2019:Data:Mining:Ethereum:Fraud:Detection, Al:2020:Labeled:Transactions:Dataset:Ethereum} and the NSL-KDD~\cite{Bala:2019:KDD:Cup:99:Dataset} datasets for blockchain-enabled IoT networks. They verified the efficiency in both blockchain and IoT network sides using a variety of evaluation metrics. 

Dey~\cite{Dey:2023:Hybrid:Meta:Heuristic:Feature:Selection:Attack:Detection} proposed a hybrid feature selection scheme combining statistical filter approaches including $\chi^2$, Pearson's correlation coefficient, and mutual information combined with a non-dominant sorting GA (NSGA-II) metaheuristic for optimizing feature selection. Specifically, filter selection methods are used to rank features based on their importance and subsequent initialization of NSGA-II allowing faster convergence to the solution. NSGA-II belongs to the evolutionary algorithms class focusing on selection, crossover, and mutation steps. Resulting populations are sorted from top to bottom without losing reasonable solutions. After selecting the most relevant features for the classification task, a SVM classifier is used in the analysis. The proposed framework is tested from a publicly available network traffic dataset (ToN-IoT) collected at a large-scale and realistic network from the Cyber Range and IoT Labs at the School of Engineering and Information Technology at UNSW Camberra~\cite{Booij:2021:TON:IoT:Dataset}. ToN-IoT covers a variety of cyberattacks inclusing ransomware, DoS, and DDoS. The proposed method reached optimal performance (99.48$\%$ accuracy) with only 13 features. 

Yazdinejad et al.~\cite{Yazdinejad:2023:Ensemble:DL:Threat:Hunting:IoT} proposed an ensemble DL model that combines LSTM and AEs to detect anomalous activities in IIoT. Output data is classified as normal or abnormal via a DT after inspecting the reconstruction error between the input and the output layer of the LSTM AE. The proposed model is evaluated in two real IoT datasets, i.e., gas pipeline and secure water treatment, which are imbalanced and have temporal dependencies. The proposed framework outperforms conventional classifiers. Despite being a promising approach for detecting anomalies in time series data, the use of LSTM requires significantly higher training than simpler models. 

Jullian et al.~\cite{Jullian:2023:DL:Cyberattacks:IoT:Networks:Distributed} implemented a distributed framework based on DL to detect different source of vulnerabilities in the IoT. Their approach consists of mainly four stages: data treatment and preprocessing, DL model training and testing, distributed framework deployment, and attack detection and classification. Their proposed approach was tested on two different datasets: the BoT-IoT dataset addresing specific attacks of IoT environments and the NSL-KDD dataset to broaden the types of cyberattacks. Due to huge imbalances in the proportion of attack vs. benign samples, they applied undersampling based on the large number of records. A standard normalization procedure is applied to both datasets to prevent models overfitting. Once the datasets are preprocessed, they evaluated two different models: a feed forward NN (FFNN) and a LSTM. Their proposed approach used a federated learning architecture to train models in a distributed fashion relying on the communication between fog and central servers. To prevent overfitting during the training procedure, they ran an optimization procedure to select the best combination of hyperparameters to achieve best evaluation metrics inclusing accuracy, precision, and recall. Their proposed distributed framework was found effective in different types of attacks achieving an accuracy of up to $99.95 \%$ across the different setups.

Basati and Faghih~\cite{Basati:2023:APAE:IoT:IDS:Symmetric:Parallel:Autoecoder} introduced an intelligent IDS framework based on an asymmetric parallel AE able to detect various attacks in IoT networks. In their preprocessing stage, they transformed 1D traffic feature vectors into 2D feature vectors with equal width and height. To extract features from more distant neighbors and association among long-range features they integrate dilatated convolution~\cite{Yu:2015:Multi:Scale:Dilatated:Convolutions} with self-attention~\cite{Vaswani:2017:Attention:is:all:You:Need}. Their approach called APAE contains two asymmetric AEs in parallel, each of them containing three successive layers of convolutional filters. APAE was evaluated in three popular public datasets named UNSW-NB15, CICIDS2017, and KDDCup99. Results showed that the proposed framework offers superior results than the state of the art. 

Gupta et al.~\cite{Gupta:2023:Integration:Digital:Twin:FL:VIoT} developed a hierarchical federated learning (HFL) anomaly detection approach to address security and data privacy concerns in the context of vehicular IoT (V-IoT). By creating a digital twin of an intelligent transportation environment, they leveraged a comprehensive virtual replica for detecting malicious activities using an anomaly detection model. To further expand the capabilities of federated learning (FL) for multi use scenarios, they developed a FL approach that allows the aggregation of gradients at multiple levels enabling the participation of multiple entities. Their proposed framework has six phases: (1) initial phase, where smart vehicle data collection begins; (2)  functional phase, where supplementary data from the external environment provide context; (3) analytic phase, where a digital twin is created for each vehicle in the V-IoT and data mining algorithms are applied to previously collected data; (4) identifying anomaly phase, where anomaly detection algorithms based on LSTM are trained to distinguish between normal and abnormal patterns; (5) collaborative phase, where the output of multiple anomaly detection algorithms is processed to improve evaluation metrics; and (6) reporting and decision phase, where anomalies are reported to relevant stakeholders.

\subsubsection{Unsupervised Learning} \label{subsubsec:IoT Unsupervised Learning}

Yasaey et al.~\cite{Yasaei:2020:IoT:CAD:Context:Aware:Anomaly:Detection:Through:Sensor:Association} proposed an adaptive context-aware anomaly detection method for securing IoT sensor data intended to run on a fog computing platform. Their approach is based on a sensor association algorithm that generates fingerprint of sensors and then cluster them to extract the context of the system. Context generation is completed by extracting the binary fingerprinting sensor values. As sensors that are affected by an event are expected to have similar patterns in their fingerprints, a clustering algorithm based on minimizing the distance between binary codifications and the Hamming distance is used. By relying on contextual information, they used a LSTM neural network and a Gaussian consensus estimator to identify the source of anomalies. In particular, for each cluster of sensors, a LSTM neural network produces a set of predictive models for each cluster. A multivariate Gaussian estimator is used to help modeling whether reconstruction errors between the real and predicted values match with the system's normal behavior. To infer the source of the anomaly a consensus algorithms checks the consistency of sensor behaviors within clusters and across clusters. Finally, to adapt the inference models with respect to concept drift, their proposed approach allows two levels of update: complete (where all the modules are retrained in order) and partial (which only retrains the predictor model). The adaptation decision is triggered based on changes in the distribution of the data. They tested their approach on the environmental training center waste water plant in Riccione~\cite{Giannoni:2018:Anomaly:Detection:IoT:Data} with synthetic anomalies. 

Gyamfi and Jarcut et al.~\cite{Gyamfi:2022:Novel:Online:IDS:IoT:OI-SVDD:AS-ELM} proposed a lightweight network intrusion detection system (NIDS) to secure industrial IoT (I-IoT) relying on an online incremental support vector data description (OI-SVDD) anomaly detection on the IIoT devices and an adaptive sequential extreme learning machine (AS-ELM) on a multiaccess edge computing (MEC) server. In their design, the OI-SVDD model is placed on the IIoT device and the AS-ELM model is placed on the MEC server at the edge network to perform deep network attack duties. Their OI-SVDD is based on using an incremental learning scheme to add samples to the training function process consisting only of support vectors at each stage. During AS-ELM, a training dataset is utilized during the initialization phase and then testing data is processed chunk by chunk. Authors tested their proposed NIDS on the UNSW-NB15 dataset and a self generated dataset showing effective performance for detecting intrusions in a realistic IIoT environment.

Li et al.~\cite{Li:2022:Few:Shot:IoT:Attack:Detection:Unsupervised:Domain:Adaptive:Regularization} proposed an adversarial unsupervised domain-adaptive regularization based on a recursive feature pyramid CNN (RFP-CNN) to detect IoT attacks more effectively. As a first step, they developed a recursive feature pyramid network (FPN) architecture~\cite{Arrington:2016:Behavioral:Modeling:IDS:Via:Immunity:Inspired:Algorithms} to extract high-level features of the RFP-CNN. The FPN focused on a hierarchical approach to predict attack features from multiple scales. Subsequently, to enable transfer learning from fewer attack samples, they developed an unsupervised domain adaptive regularization model. They showed the potential of their approach for anti-noise performance and short running time. Their proposed framework is validated using four intrusion detection datasets used in the IoT space. In doing so, they constructed few shots datasets from the ICSX2012FS and CICIDS2017FS datasets sampling a few attack samples.

Table~\ref{table: Adaptive anomaly detection in IoT} summarizes key attributes of the papers reviewed in Section~\ref{subsec:IoT}.
\begin{table*}[htp!]
\centering
\caption{Summary of AAD methods in IoT applications.}
\label{table: Adaptive anomaly detection in IoT}
\small 
\begin{adjustbox}{max width=\textwidth}
\begin{tabular}{|p{2.5cm}|p{1.5cm}|p{1.5cm}|p{2cm}|p{2.5cm}|p{3.5cm}|p{3.5cm}|}
\hline
{Reference} & {Model} & {Algorithm} & {Dataset} & {Attack} & {Strengths} & {Weaknesses} \\
\hline
Li et al.~\cite{Li:2019:System:Statistics:Learning:IoT} 2019 & Supervised & LR, NN, RNN & Simulated IoT network & Unathorized access, port scan, virus, flood & Online learning capability with adaptive threshold & Tested on simple DoS attacks \\

Yasaei et al.~\cite{Yasaei:2020:IoT:CAD:Context:Aware:Anomaly:Detection:Through:Sensor:Association} 2020 & Unsupervised & LSTM, Multivariate Gaussian estimator & IoT testbed & Simulated data injection attacks & LSTM computational complexity & Lack of continual learning \\

Gopalakrishnan et al.~\cite{Gopalakrishnan:2020:DL:Data:Offloading:Mobile:Edge:Computing:System} 2020 & Supervised & BiLSTM, ASCE, DBN & Simulated traffic data & Not specified & Traffic prediction, data offloading, and cyberattack detection & Lack of continual learning \\

Bibi et al.~\cite{Bibi:2022:Deep:AI:Cyber:Threat:Analysis:IIoT} 2022 & Supervised & ConvLSTM2D  & Network traffic & MitM, DoS, botnet malware, and Recon & Catch sophisticated threats and attacks & Lack of continual learning \\

Li et al.~\cite{Li:2022:Few:Shot:IoT:Attack:Detection:Unsupervised:Domain:Adaptive:Regularization} 2022 & Unsupervised & RFP-CNN & Network traffic & Networking attacks & Quick inference & Lack of continual learning \\

Gyamfi and Jurcut~\cite{Gyamfi:2022:Novel:Online:IDS:IoT:OI-SVDD:AS-ELM} 2022 & Unsupervised & OI-SVDD and AS-ELM & Network traffic & Networking attacks & Online learning capability & Rely on a MEC server to operate effectively \\

Dey et al.~\cite{Dey:2023:Hybrid:Meta:Heuristic:Feature:Selection:Attack:Detection} 2023 & Supervised & NSGA-II & Network traffic & Ransomware, DoS, and DDoS & Effective feature selection & Lack of continual learning \\

Yazdinejad et al.~\cite{Yazdinejad:2023:Ensemble:DL:Threat:Hunting:IoT} 2023 & Supervised & LSTM-AE & IIoT traffic data & Network operation under attack & Temporal dependencies modeling & Lack of continual learning  \\

Yazdinejad et al.~\cite{Yazdinejad:2023:Secure:Fuzzy:Blockchain:Framework} 2023 & Supervised & Fuzzy DL & Blockchain and IoT & Phishing and Ethereum fraudulent transactions & Tight integration between control/data planes & Lack of continual learning \\

Jullian et al.~\cite{Jullian:2023:DL:Cyberattacks:IoT:Networks:Distributed} 2023 & Supervised & FCNN and LSTM & IoT network data & DoS, DDoS, Keylogging, Data Theft & Distributed environment & Lack of continual learning \\

Basati and Faghih~\cite{Basati:2023:APAE:IoT:IDS:Symmetric:Parallel:Autoecoder} 2023 & Supervised & Asymmetric parallel autoencoder & Network data & Networking attacks & Long-range feature extraction & Lack of continual learning \\

Gupta et al.~\cite{Gupta:2023:Integration:Digital:Twin:FL:VIoT} 2023 & Supervised & HFL and LSTM & Case scenario & Not specified & Privacy preservation through FL & Lack of continual learning \\
\hline
\end{tabular}
\end{adjustbox}
\end{table*}

\subsection{AAD in smart grids} \label{subsec:SG}

\subsubsection{Supervised Learning} \label{subsubsec:smart grid Supervised Learning}

Adhikari et al.~\cite{Adhikari:2017:Applying:Adaptive:Threes:Real:Time:IDS} proposed a cyber-power event and intrusion detection system (EIDS) that can be used for multiclass or binary classification of traditional power system contingencies and cyber-attacks. In doing so, they process continuous streams of high speed data from wide area monitoring systems (WAMS) and used a HAT augmented with the drift detection method (DDM) and adaptive windowing (ADWIN) that classifies traffic in real-time. HAT data stream mining is based on the idea of concept adapting very fast DTs (CVFDTs). In particular, HAT creates DTs from the data stream and updates the three after inspecting each sample. HAT does not require samples to be stored in memory as nodes in the tree hold holds rich information to perform classification. DDM compared the statistics of two windows to detect concept drifts when the number of errors increases beyond a threshold. ADWIN is a parameter free adaptive size sliding window technique that is used to detect concept change and trigger model revision. DDM and ADWIN together enhances HAT's ability to detect change and update the model. Experiments on a wide area measurement system with hardware in the loop testbed with simulated attacks showed that the combined approach of HAT, DDM, and ADWIN provides greater classification accuracy along with a small memory footprint and fast evaluation.


Wang and Govindarasu~\cite{Wang:2020:Multi:Agent:Attack:Resilient:Smart:Grid} presented a data-driven anomaly detection and adaptive load rejection scheme within a decentralized system integrity protection (SIP) for smart grids. In doing so, they leveraged a SVM embedded layered DT (SVMLDT) that arrives at a decision based on the consensus among all interconnected agents. The SVMLDT works by segregating the training dataset into subsets based on all nominal features and then reducing the dimensionality of the feature space. Then SVMLDT applies a DT based SVM (DTSVM) for supervised classification. The proposed framework responds adaptively to DoS attacks by separating the multi-agent system into several interconnected subgroups so that within one subgroup, the real-time load profiles can be still shared globally. They used a real load rejection SIP scheme adopted by salt-river project to fir in the IEEE 39-bus model as a study case. Their results show that the proposed SIP can effectively detect anomalous grid operation states and then adjust its actions accordingly to adapt to the under attack situations. 

Camana et al.~\cite{Camana:2020:Randomized:Threes:Stealthy:Cyberattack:SG} proposed an approach for cyberattack detection in smart grids based on an extremely randomized three (Extra-Trees) algorithm and kernel principal component analysis (K-PCA) for dimensionality reduction. Extra-Trees uses a large number of DTs and chooses a split rule based on a random subset features and a partially random cut point. K-PCA is used to tackle the increasing computational complexity of big power systems by considering non-linearities inherent in data with complicated structures. They studied the attack detection problem where the labels of the samples are randomly corrupted. Specifically, a percentage of true labels in the training is flipped also know as label noise. They tested their proposed approach on the IEEE 57-bus and 118-bus systems. Their numerical results show that their approach outperforms state of the art approaches. 


Liao et al.~\cite{Liao:2022:Divergence:Transferability:Analysis:Adaptive:Smart:Grid:IDS} proposed a divergence-based transferability analysis to decide whether or not to apply transfer learning and automatically adapt a smart grid intrusion detection strategy. In particular, they explored three metrics to capture the similarity of data distributions to understand the relationship between detector's accuracy drop and similarity. Following up on this analysis, they trained two regression models to approximate the similarity and accuracy relationship needed to predict accuracy drops, which indicates the need for transfer learning. A domain adversarial neural network (DANN) classifier is adopted as transfer learning model. To validate their effectiveness, they used datasets from real normal operations from ISO New England~\cite{Muzhikyan:2019:SOARES:Dataset} and simulated attacks from the IEEE 30-bus system in different conditions including attack timing, location, and both. Ultimately, their approach shows that the DANN can be timely triggered to achieve an accuracy improvement over $5.0\%$. 

\subsubsection{Unsupervised Learning} \label{subsubsec:smart grid Unsupervised Learning}

Li et al.~\cite{Li:2016:Dirichlet:Based:Detection:Smart:Grid} proposed a Dirichlet-based probabilistic model to asses the reputation levels of decentralized local agents (LA). Initial reputation levels of LAs used historical data to train the proposed model. To detect opportunistic attackers, they develop an adaptive detection algorithm with a reputation incentive mechanism. Specifically, they used the Bayes rule to assist the control center in making informed decisions about LAs being compromised. In doing so, they used a Dirichlet distribution as a prior distribution. Initial beliefs combined with a series of historical observations shape the posterior distribution that is best suited for the reputation model. To estimate the overall status of combined LA's behaviors, they leverage a reputation level in their scheme based on the graded mean value of each compliance level. They demonstrated the utility of the proposed framework using data from IEEE-39 power system using the PowerWorld simulator.

\subsubsection{Reinforcement Learning} \label{subsubsec:smart grid reinforcement Learning}

Hu et al.~\cite{Hu:2022:RL:Adaptive:Feature:Boosting:SG} proposed a RL-based adaptive feature boosting that leverages a series of AEs capturing critical features from multi-source smart grid data for the classification of normal, fault, and attack events. They used multiple AEs to extract representative features from different feature sets which are extracted through a weighted feature sampling process. These extracted features are informed by a reinforcement learning approach called deep deterministic policy gradient (DDPG) to determine the feature sampling probability based on classification accuracy. AE-based extracted features are then feed in a RF classifier as base classifier along with an ensemble to distinguish between different types of incidents. They evaluated their proposed approach in two realistic datasets collected from hardware-in-the-loop (HIL) and WUSTIL-IIOT-2021 security testbeds showing an increase in classification accuracy with respect to the vanilla adaptive feature boosting.

Table~\ref{table: Adaptive anomaly detection in smart grids} summarizes key attributes of the papers reviewed in Section~\ref{subsec:SG}.
\begin{table*}[htp!]
\centering
\caption{Summary of AAD methods in smart grid applications.}
\label{table: Adaptive anomaly detection in smart grids}
\small 
\begin{adjustbox}{max width=\textwidth}
\begin{tabular}{|p{2.5cm}|p{1.5cm}|p{1.5cm}|p{2cm}|p{2.5cm}|p{3.5cm}|p{3.5cm}|}
\hline
{Reference} & {Model} & {Algorithm} & {Dataset} & {Attack} & {Strengths} & {Weaknesses} \\
\hline
Li et al.~\cite{Li:2016:Dirichlet:Based:Detection:Smart:Grid} 2016 & Unsupervised & Dirichlet-based probabilistic model & IEEE 39-bus simulations & FDIA & Continual learning & Test on simulated data  \\

Adhikari et al.~\cite{Adhikari:2017:Applying:Adaptive:Threes:Real:Time:IDS} 2017 & Supervised & HAT, DDM, ADWIN & HIL testbed simulation & fault, line maintenance, load fluctuation, and cyber-attacks scenarios & Continual learning & Need of labeled data \\

Wang and Govindarasu~\cite{Wang:2020:Multi:Agent:Attack:Resilient:Smart:Grid} 2020 & Supervised & SVMLDT & IEEE 39-bus simulation & DoS and Replay attacks & Continual learning & Computational complexity \\

Camana et al.~\cite{Camana:2020:Randomized:Threes:Stealthy:Cyberattack:SG} 2020 & Supervised & Extra-Threes and K-PCA & IEEE 118-bus and 57-bus simulations & FDIA & Dimensionality reduction reduces complexity of models & Lack of continual learning \\

Liao et al.~\cite{Liao:2022:Divergence:Transferability:Analysis:Adaptive:Smart:Grid:IDS} 2022 & Supervised & DANN & Public load demand & FDIA & Continual learning & Framework complexity \\

Hu et al.~\cite{Hu:2022:RL:Adaptive:Feature:Boosting:SG} 2022 & Reinforcement learning & AEs and DDPG & HIL and WUSTIL-IIOT-202 & Line maintenance, data injection, remote tripping command, relay setting change & Continual learning & Tested on small-scale testbed \\
\hline
\end{tabular}
\end{adjustbox}
\end{table*}

\section{Discussion} \label{sec:discussion}


This SLR focuses on the use of AAD methods for CPS. We classify the review studies using a new taxonomy based on CPS applications and ML algorithms. This section covers the SLR findings, current approach limitations, and future research directions in AAD for CPS.

\subsection{Findings} \label{subsec:findings}

Supervised, unsupervised, and reinforcement learning can be trained in both offline and online fashion on traditional ML, DL, and hybrid models. Unsupervised approaches are better equipped to detect unknown anomalies than supervised learning algorithms. Most unsupervised methods rely only on benign data (also know as one-class classification~\cite{ Chandola:2009:Anomaly:Detection:Survey, Yuan:2022:Trustworthy:AD:Survey}). Semi-supervised anomaly detection typically assumes a small number of labelled normal and abnormal samples and a large number of unlabeled samples in the training dataset. Therefore, we group them under the unsupervised category. We notice that a few of the reviewed papers proposed semi-supervised approaches, where the training data consist only from normal data without anomalies~\cite{Goldstein:2016:Comparative:Evaluation:Unsupervised:Anomaly:Detection}. RL addresses the challenge of learning from interaction with an environment to achieve long term goals of protecting CPS against cyberattacks. Although RL is a promising approach, we noticed that it does not have widespread use in practice~\cite{Nguyen:2021:DRL:Cybersecurity}.

As normal streaming data in CPS is easier to collect than attack data, unsupervised learning is a promising approach for AAD in CPS. We observe that OCSVM (a traditional ML model) and AEs (DL based) were commonly used approaches in unsupervised learning. We notice that a combination of sequential models based on RNNs such as LSTM and GRU with other DL architectures including CNNs or rule-based models have been effective on increasing detection capabilities for sophisticated attacks. As they tend to be a good fit to model the normal temporal dependencies in time series data generated by CPS, they have been used successfully in unsupervised fashion. Unsupervised models can detect a wider range of attacks including previously unseen attacks compared to supervised models at the expense of producing higher false positive rates. Generally, DL methods have achieved better performance than traditional ML methods. Nonetheless, due to expensive resource requirements and high latency, DL methods might not be the best option to deploy AAD solutions for CPS, given their usual limited resource availability. We notice that the use of hybrid and ensemble models can help to overcome the limitations of individual models by increasing the robustness of decision making. We observe that only two reviewed papers use RL strategies, suggesting there is plenty of room for exploration and promise to detect novel attacks in streaming environments. 



\begin{figure*}[p]
\centering
\includegraphics[width=0.78\textwidth]{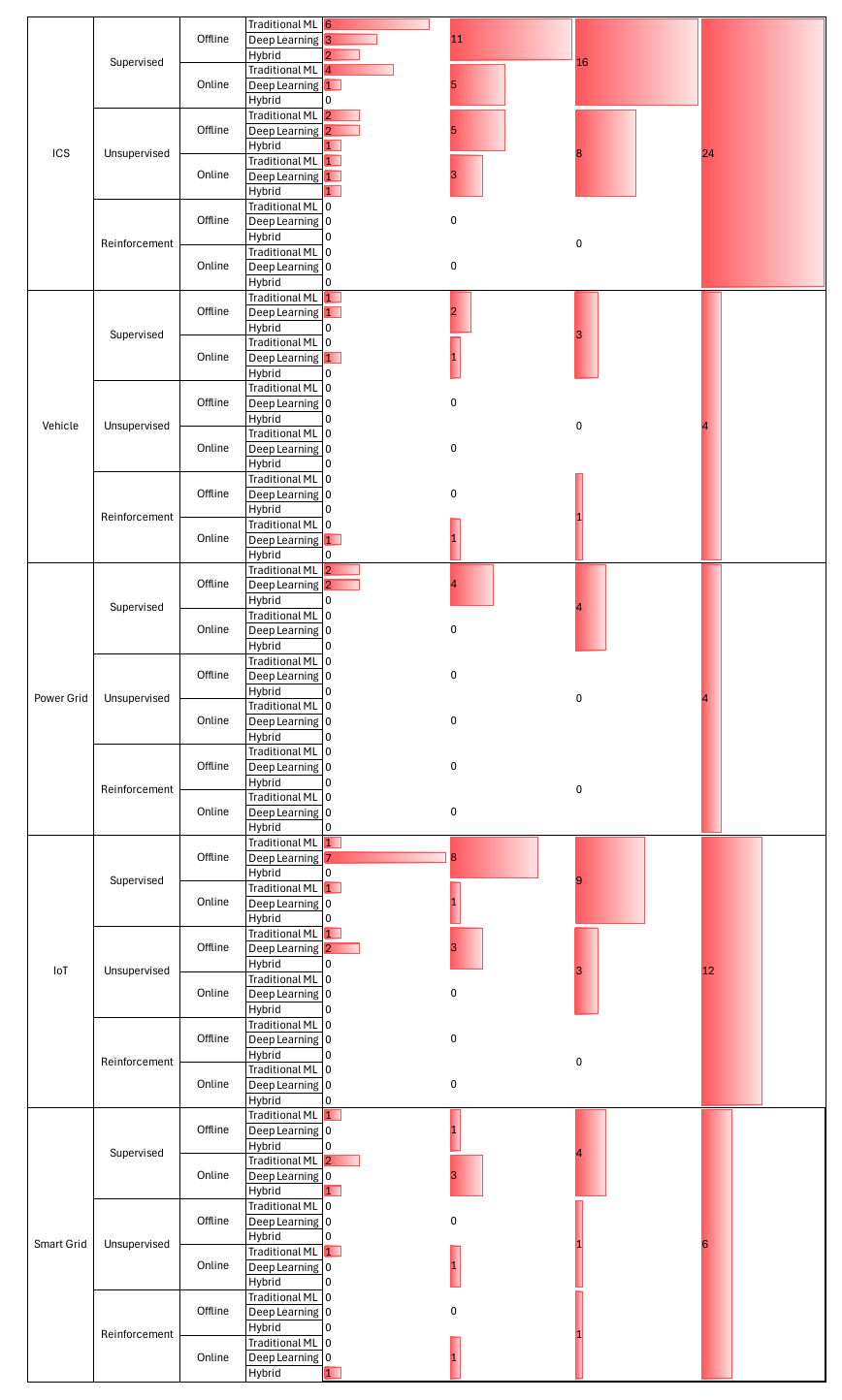}
\caption{AAD for CPS distribution among application, learning paradigm, data management, and algorithm categories. }
\label{fig:bar plots}
\end{figure*}

We apply our taxonomy to the reviewed work in Figure~\ref{fig:bar plots}. It reveals how the reviewed AAD for CPS work distributes across CPS applications, learning paradigm, data management, and algorithms. We make the following observations: (1) most of the reviewed work focused on ICS; this corresponds to more than 50$\%$ (24 out of 47) of the reviewed papers; (2) the vast majority of proposed literature focused on supervised models that require labels; this corresponds to three quarters (36 out of 47) of the reviewed papers; (3) most proposed methods are tested in an offline setting; this corresponds to nearly two-thirds of the reviewed papers (32 out of 47); (4) most of the proposed methods rely on DL techniques; this correspond to nearly $50\%$ (21 out of 47) of the reviewed papers. One of the most prominent findings of our analysis is the lack of online evaluation of the proposed methods with only about one third (16 out of 47) of the reviewed papers discussing it.
Given the complexity of attacks and the conditions of the deployment environment, AAD methods are expected to use diverse techniques within resource-constrained settings. Our SLR suggests that an online unsupervised ensemble method is the most suitable approach for AAD in CPS, given the required constraints.


\subsection{Limitations} \label{subsec:limitations}


\subsubsection{Paper Retrieval Omissions} \label{subsubsec: Paper Retrieval Omissions}

Our SLR may potentially overlook relevant papers during the search process. While collecting AAD research papers from various publishers, there is a risk of missing those with incomplete abstracts or without adaptation focus or the CPS keyword. To address this, we used a systematic approach that combined manual searching, automatic searching, and forward and backward search to minimize the chance of missing relevant papers. We searched for AAD papers in CPS from leading engineering and science publishers that sponsor top conferences in data mining and computer security, extracting AAD and CPS keywords for manual searches. We also conducted automatic searches using a carefully selected set of keywords. To further expand our results, we applied forward and backward search snowballing techniques.

\subsubsection{Paper Selection Bias} \label{subsubsec: Paper Selection Bias}

Manual inspection of papers has inherent limitations and potential biases. We selected papers using a process that combined manual and automated steps, followed by validation based on quality criteria. However, the manual validation stage may be biased by the researcher's judgment, affecting the accuracy of paper quality assessment. To address this, the coauthors---experts in data mining, anomaly detection, and cybersecurity---conducted an additional in-depth review for relevance and quality screen (see Figure~\ref{fig:des-dia}). This step aims to improve the accuracy of paper selection and reduce the risk of omissions and misclassifications. By implementing these measures, we aim to ensure the accuracy and integrity of the selected papers, minimize selection bias, and enhance the reliability of our SLR.

\subsection{Future Research Directions} \label{subsec:future research directions}


\subsubsection{AAD Datasets} \label{subsub:AAD Datasets}

Evaluation of AAD methods in CPS is highly dependent on the data being used. The use of low quality data with absent sophisticated attacks on CPS may lead to biased and incorrect conclusions. Some datasets are based on injecting simulated attacks under realistic conditions, which hinders nuances of CPS. Thus, it is difficult to evaluate, compare, and improve AAD without having proper datasets. Associated reasons for this limitation include: (1) cost to produce real attacks in CPS, (2) associated risk for introducing attacks in CPS, (3) the disclosure of private information~\cite{verma2024comprehensive}. Looking forward, the creation of CPS attack datasets in controlled but realistic conditions and the combination of multiple existing datasets is an interesting direction in the future. A good example is the ROAD dataset~\cite{verma2024comprehensive} in the vehicle domain but other datasets are needed in other CPS applications. 

\subsubsection{Low Detection Latency} \label{subsub:Detection Latency}

As CPS transmit data in real-time, AAD techniques should act rapidly to take appropriate countermeasures in near real-time. However, most of the studies that we reviewed proposed solutions incapable of detecting attacks in near real-time. In addition, the DL-based methods we reviewed assume availability of a large number of computational resources in the cloud to arrive at conclusions. However, since CPS have multiple components, connection stability is a key factor for the deployment in cloud environments. Further AAD methods that are able to effectively process streaming data and arrive at detection decisions with low latency is a needed direction of study in the future.     

\subsubsection{Consistent Evaluation Metrics} \label{subsub:Consistent Evaluation Metrics}

AAD reviewed work performed evaluation tests of their approaches on datasets from different nature, including real and synthetic data. Evaluation metric comparisons using real and synthetic data are possible thanks to the common benign and attack data usually coming with the datasets. However, we notice a variety of evaluation metrics usually reported in the revised studies, so that studies do not report common metrics that allows a head-to-head comparison. Common evaluation metrics reported include accuracy, precision, recall, F1 score, and AUC-ROC. In the context of anomaly detection, it is common to face imbalanced datasets, making accuracy an inappropriate metric in the area. Recent studies have suggested the use of the Matthews correlation coefficient (MCC) which in general is intended to address the unbalanced data issues commonly found in anomaly detection datasets~\cite{Chicco:2020:MCC:Over:F1}. Therefore, to make fair comparisons on AAD approaches suggested metrics include precision, recall, false positive rate, false negative rate, and MCC. In addition, detection latency found in a handful of studies should be also reported particularly in the streaming context. Thus, we expect future work including suggested metrics as part of their evaluation criteria. 

\subsubsection{Unsupervised AAD}\label{subsub:Unsupervised AAD}

Since attack datasets are scarce and labeling is difficult and costly, unsupervised learning techniques (e.g., clustering, OCSVM, AEs) are well suited for AAD. The fundamental assumption in unsupervised learning is that only benign data is used to model normal behavior and a threshold is used as criteria for decision making. With abundant benign data in streaming scenarios, an important future research direction is to develop adaptive threshold mechanisms that can be updated online and respond to concept drifts.

\subsubsection{DL-Based AAD Requires Abundant Data} \label{subsub:DL-based AAD Requires Abundant Data}

Since many AAD approaches rely on DL, the lack of realistic attack and benign datasets is a common challenge. Therefore learning from a few samples in a dynamic, changing environment is key. Other fields such as computer vision and natural language processing have developed powerful machinery to learn from limited data, including transfer learning~\cite{Pan:2009:Transfer:Learning:Survey}, one-shot learning~\cite{Vinyals:2016:Matching:Networks:One:Shot:Learning}, and zero-shot learning~\cite{Xian:2017:Zero:Shot:Learning:Good:Bad:Ugly}. Building on previous work and adapting it to use small labeled datasets for AAD is an important future research direction.

\subsubsection{Model's Complexity} \label{subsub:Model's Complexity}

The reviewed papers focused mainly on the algorithmic and software aspects of their proposed AAD techniques. However, the deployment of these techniques is often overlooked due to the hardware constraints common in the CPS applications we studied. Since host-based deployment is unfeasible due to additional security and privacy related concerns, a different approach is needed so that the online conditions needed to deploy AAD are meet in a cost-effective manner. Therefore, increasing attention must be placed on the use and integration with edge, fog, or cloud computing infrastructures. To this, deployment and validation of AAD in edge-fog-cloud computing infrastructures can be considered an important future research direction.

\subsubsection{Adversarial AAD} \label{subsub:Adversarial AAD}

A few proposed AAD approaches are effective in identifying previously unseen anomalies in CPS with high detection rates. However, these models remain vulnerable to adversarial attacks, including white-box, black-box, and tampering attacks~\cite{Chakraborty:2021:Survey:Adversarial:Attacks:Defenses, Aloraini:2024:Adversarial:Attacks:IDS:False:Alarms}. None of the reviewed studies address how to protect the proposed AAD approaches from attacks. Therefore, developing secure AAD for CPS in adversarial settings is a promising and challenging direction for future research.~\cite{Yuan:2022:Trustworthy:AD:Survey, Mohus:2023:Adversarial:Robustness:Unsupervised:Learning}. We observe that existing solutions from the AI security domain could be adapted for AAD in CPS~\cite{Goodge:2021:Robustness:Autoencoders:AD:Adversarial:Attacks, Lo:2022:Adversarially:Robust:One:Class:Novelty:Detection}.

\subsubsection{Explainable AAD} \label{subsub:Explainable AAD}

The reviewed papers primarily focus on improving classification outcomes for AAD. Most proposed methods, particularly those based on DL, emphasize black-box anomaly detection for making critical predictions. However, various stakeholders are increasingly demanding explainability.~\cite{Preece:2018:Stakeholders:Explainable:AI}. Explainable anomaly detection refers to a model's ability to clarify why and when it identifies an anomaly~\cite{Li:2023:Survey:XAD}. Prioritizing explainable AAD is a critical research direction, as decision-making in CPS demands models that can provide clear explanations of detection results without compromising prediction quality.

\section{Conclusion} \label{sec:conclusion}

We have performed a SLR in the field of AAD in CPS. We performed our searchers in five prominent research databases (i.e., IEEE Explore, ACM Digital Library, Emerald Insight, Springer Link, and Science Direct) and execute forward and backward snowballing search to maximize the literature search coverage. After analyzing 47 research papers and 18 review articles, this study introduces a novel taxonomy based on attacks, CPS applications, learning paradigm (i.e., supervised, unsupervised, and reinforcement learning), and ML algorithms. In particular, we outline algorithms, datasets, attack characteristics, strengths and weaknesses developed in each of the papers. Details about the learning paradigm used are discussed with respect to each step guiding attack detection strategy across CPS applications.

By using a known and standardized approach, our SLR ensures that key papers in the searched databases are found. Thus, researchers and practitioners should not need to repeat this work to find relevant publications from the period 2013 to 2023 (November). Still, the SLR approach allows the work to be repeated in the future to track field developments. Our categorization provides a clear overview of AAD in CPS and related research, making it easy to find relevant papers for specific CPS applications. The number of papers in each category shows which research areas were seen as important and challenging during the studied period.

The current findings indicate that most of the previous studies investigated only a single aspect of adaptation (either data processing or model adaptation). Limited studies have provided a comprehensive overview of both aspects of AAD at the same time. Hence, this SLR contributed to the anomaly detection literature by summarizing different adaptive ML-based anomaly detection algorithms into distinct CPS applications. The intersection of these topics has not been discussed previously. The identified and categorized studies not only add the conceptual discussion in the filed of AAD but also provided several enlightened ideas for researchers and practitioners. Similarly, different types of data analysis methods can also diversify our investigation in future studies and enrich our results.

Finally, based on prior research, we highlight several directions and considerations for future studies. These directions and considerations include, tight integration between near real-time data processing and an adaptive learning mode, explainable and actionable detection predictions, and quantifying the uncertainty of detection predictions.

\section{Acknowledgements} \label{sec:acknowledgements}

This research was sponsored in part by Oak Ridge National Laboratory’s (ORNL’s) Laboratory Directed Research and Development program and by the DOE. There was no additional external funding received for this study. 
The funders had no role in study design, data collection and analysis, decision to publish, or preparation of this manuscript.

\bibliographystyle{IEEEtran}
\bibliography{ 99-references}

 

\begin{IEEEbiography}[{\includegraphics[width=1in,height=1.25in,clip,keepaspectratio]{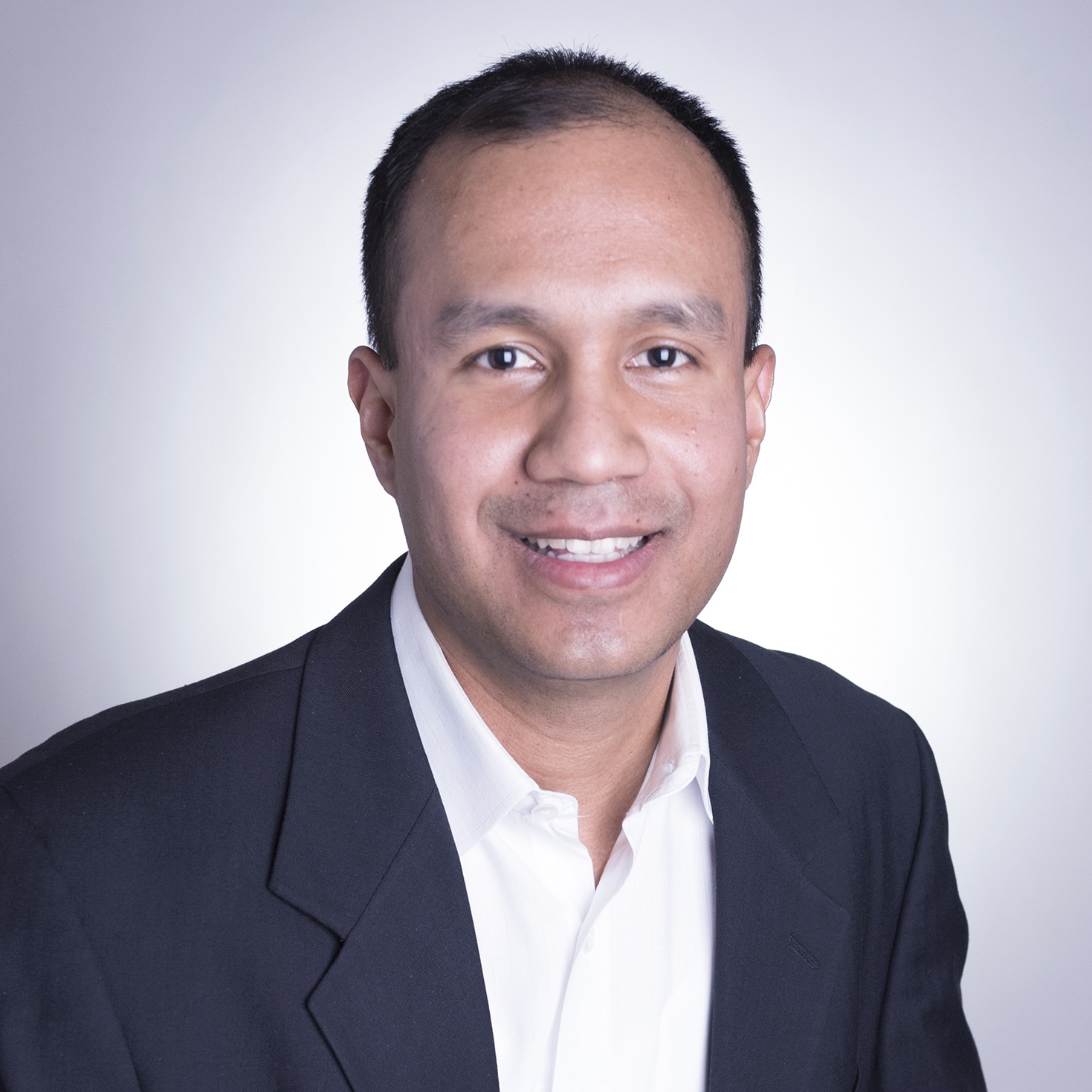}}]{Pablo Moriano}
(Senior Member, IEEE) received B.S. and M.S. degrees in electrical engineering from Pontificia Universidad Javeriana in Colombia and M.S. and Ph.D. degrees in informatics from Indiana University Bloomington, Bloomington, IN, USA. He is a research scientist with the Computer Science and Mathematics Division at Oak Ridge National Laboratory, Oak Ridge, TN, USA. His research lies at the intersection of data science,
network science, and cybersecurity. In particular, he uses data-driven and computational methods to discover, understand, and detect anomalous behavior in large-scale networked systems. Applications of his research range across multiple disciplines, including, the detection of exceptional events in social media, Internet route hijacking, insider threat behavior in version control systems, and anomaly detection in cyber-physical systems. Dr. Moriano is a member of ACM and SIAM.
\end{IEEEbiography}

\vspace{11pt}


\begin{IEEEbiography}[{\includegraphics[width=1in,height=1.25in,clip,keepaspectratio]{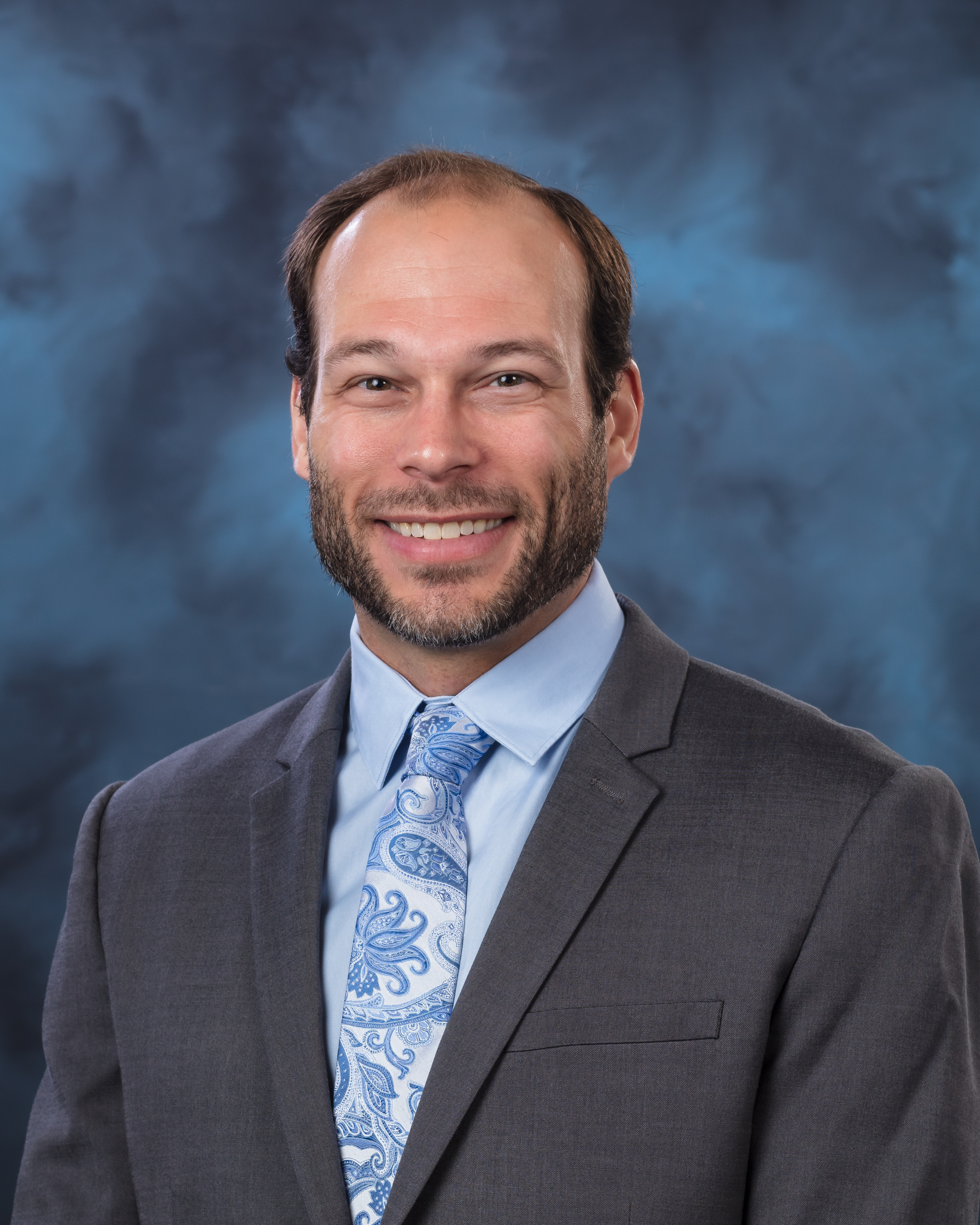}}]{Steven C. Hespeler} earned his B.S. in Engineering with a minor in Mathematics from Roger Williams University, RI, USA. Following this, he completed a M.S. in Industrial Engineering and Ph.D. in Industrial Engineering with a minor in Applied Statistics both from from New Mexico State University, NM, USA. He is a Postdoctoral Research Associate with the Computer Science and Mathematics Division, Oak Ridge National Laboratory, Oak Ridge, TN, USA. His research interests are a fusion of data science, non-destructive evaluation, and advanced signal processing. He develops innovative data-driven models using machine and deep learning techniques to address complex engineering challenges. His work spans diverse applications, including in-situ process monitoring, anomaly detection, and defect identification in manufacturing, cybersecurity, and robotic inspection.

\end{IEEEbiography}

\begin{IEEEbiography}[{\includegraphics[width=1in,height=1.25in,clip,keepaspectratio]{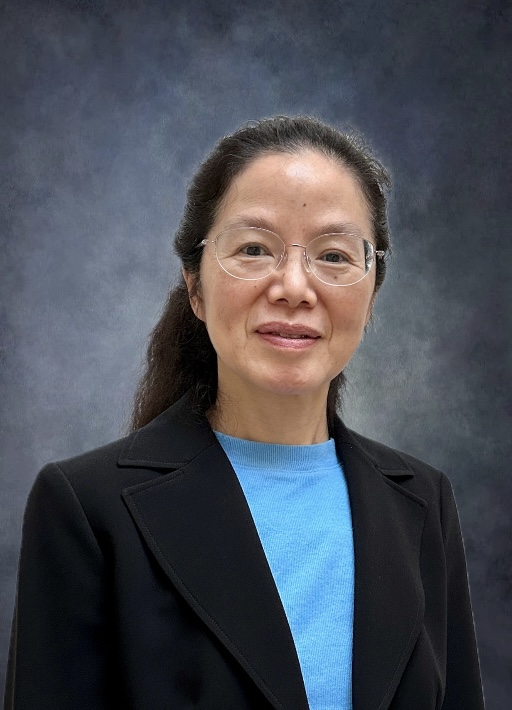}}]{Mingyan Li} is a Senior cybersecurity researcher at Oak Ridge National Laboratory (ORNL). She joined ORNL in 2021, after working at Boeing Research and Technology for 15 years. She received a Ph.D. in EE from University of Washington. Her research interests include cybersecurity data analytics, AI-based IoT/cloud system security/privacy, and key management; within and beyond distributed computing and transportation system domains.

\end{IEEEbiography}

\begin{IEEEbiography}[{\includegraphics[width=1in,height=1.25in,clip,keepaspectratio]{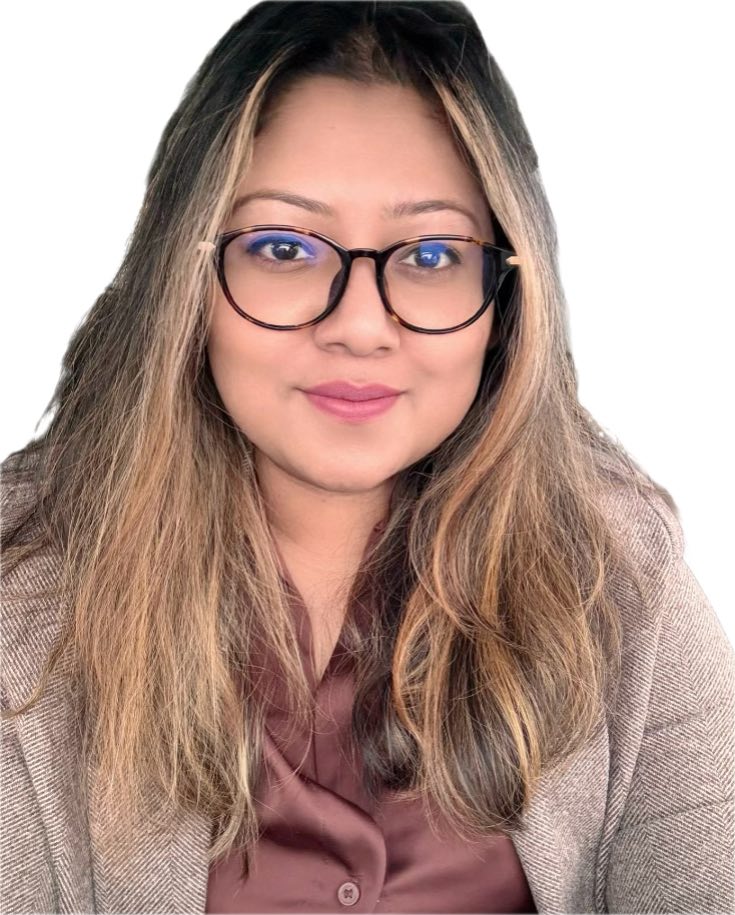}}]{Maria Mahbub} (Member, IEEE) is a Staff Researcher at Oak Ridge National Laboratory. She holds a B.S. and an M.S. in Mathematics and Applied Mathematics from the University of Dhaka, Bangladesh, and an additional M.S. and a Ph.D. in Computer Science from the University of Tennessee, Knoxville, USA. Her expertise lies in machine learning, deep learning model development, and natural language processing. Currently, she serves as the technical lead for information extraction and predictive modeling in projects funded by the U.S. Department of Veterans Affairs. Her research also focuses on enhancing the robustness of deep learning models through adversarial attack and defense strategies.

\end{IEEEbiography}

\end{document}